\documentclass[12pt]{article}

\usepackage[a4paper, total={18.5cm, 26.5cm}]{geometry}
\usepackage{authblk}
\usepackage[inline]{enumitem}
\usepackage{eurosym}
\usepackage{graphicx}
\usepackage{hyperref}

\hypersetup{
    colorlinks=true,
    linkcolor=green,
    pdftitle={FAIR Data Pipeline}
    }

\begin{document}

\title{FAIR Data Pipeline: provenance-driven data management for traceable scientific workflows}

\author[1,2]{Sonia Natalie Mitchell}
\author[3]{Andrew Lahiff}
\author[3]{Nathan Cummings}
\author[3]{Jonathan Hollocombe}
\author[4]{Bram Boskamp}
\author[5]{Ryan Field}
\author[6]{Dennis Reddyhoff}
\author[3]{Kristian Zarebski}
\author[7]{Antony Wilson}
\author[3]{Bruno Viola}
\author[4]{Martin Burke}
\author[8]{Blair Archibald}
\author[9]{Paul Bessell}
\author[10]{Richard Blackwell}
\author[9]{Lisa A Boden}
\author[3]{Alys Brett}
\author{Sam Brett}
\author[5]{Ruth Dundas}
\author[8,2]{Jessica Enright}
\author[7]{Alejandra N. Gonzalez-Beltran}
\author[4,2]{Claire Harris}
\author[11]{Ian Hinder}
\author[10]{Christopher David Hughes}
\author[4]{Martin Knight}
\author[10]{Vino Mano}
\author[5,2]{Ciaran McMonagle}
\author[12,2]{Dominic Mellor}
\author[1,2]{Sibylle Mohr}
\author[4,2]{Glenn Marion}
\author[1,2]{Louise Matthews}
\author[4,2]{Iain J. McKendrick}
\author[4]{Christopher Mark Pooley}
\author[13]{Thibaud Porphyre}
\author[14]{Aaron Reeves}
\author[19]{Edward Townsend}
\author[6]{Robert Turner}
\author[15]{Jeremy Walton}
\author[1,2]{Richard Reeve}

\affil[1]{Institute of Biodiversity, Animal Health and Comparative Medicine, College of Medical, Veterinary and Life Sciences, University of Glasgow, Glasgow, G12 8QQ, UK}
\affil[2]{Boyd Orr Centre for Population and Ecosystem Health,
University of Glasgow, Glasgow, UK}
\affil[3]{United Kingdom Atomic Energy Authority, Didcot, OX14 3DB, UK}
\affil[4]{Biomathematics and Statistics Scotland (BioSS), James Clerk Maxwell Building, Peter Guthrie Tait Road, The King's Buildings, Edinburgh, EH9 3FD, UK}
\affil[5]{MRC/CSO Social and Public Health Sciences Unit, Institute of Health and Wellbeing, College of Medical, Veterinary and Life Sciences, University of Glasgow, Glasgow, G12 8QQ, UK}
\affil[6]{Department of Computer Science, University of Sheffield, Regent Court, Sheffield, S1 4DP, UK}
\affil[7]{Science and Technology Facilities Council, Harwell Campus, UK}
\affil[8]{School of Computing Science, College of Science and Engineering, University of Glasgow, Glasgow, G12 8QQ, UK}
\affil[9]{Roslin Institute, University of Edinburgh, Edinburgh, UK}
\affil[10]{Man Group plc, Riverbank House, 2 Swan Lane, London, EC4R 3AD,  UK}
\affil[11]{The University of Manchester, Research IT, Manchester, M1 3BU, UK}
\affil[12]{School of Veterinary Medicine, College of Medical, Veterinary and Life Sciences, University of Glasgow, Glasgow, G61 1QH, UK}
\affil[13]{VetAgro Sup, UMR5558 Laboratoire de Biométrie et Biologie Évolutive, Campus vétérinaire de Lyon, Marcy-l'Etoile, FR 69280}
\affil[14]{Scotland's Rural College (SRUC), Peter Wilson Building, The King's Buildings, West Mains Road, Edinburgh, EH9 3JG, UK}
\affil[15]{UK Earth System Model Core Group, Met Office, Exeter, EX1 3PB, UK}
\date{}


\begin{titlepage}
    \maketitle
\end{titlepage}


\newgeometry{total={16cm, 24cm}}




\begin{abstract}
Modern epidemiological analyses to understand and combat the spread of disease depend critically on access to, and use of, data. Rapidly evolving data, such as data streams changing during a disease outbreak, are particularly challenging. Data management is further complicated by data being imprecisely identified when used.

Public trust in policy decisions resulting from such analyses is easily damaged and is often low, with cynicism arising where claims of ``following the science'' are made without accompanying evidence. Tracing the provenance of such decisions back through open software to primary data would clarify this evidence, enhancing the transparency of the decision-making process.
    
Here, we demonstrate a Findable, Accessible, Interoperable and Reusable (FAIR) data pipeline. Although developed during the COVID-19 pandemic, it allows easy annotation of any data as they are consumed by analyses, or conversely traces the provenance of scientific outputs back through the analytical or modelling source code to primary data. Such a tool provides a mechanism for the public, and fellow scientists, to better assess scientific evidence by inspecting its provenance, while allowing scientists to support policy-makers in openly justifying their decisions. We believe that such tools should be promoted for use across all areas of policy-facing research.
\end{abstract}

\section{Introduction}

Historically, models and analyses used to support advice to government have not been publicly available as public policies are implemented.  Typically, some materials would only subsequently become public via traditional publication routes, with the delays that this implies. Technological advances and increasingly influential ideas from open source and reproducible science mean this approach is no longer tenable. During the current COVID-19 pandemic, many models used by the Scientific Pandemic Influenza Group on Modelling (SPI-M), who advise the United Kingdom Government on human infectious disease threats based on infectious disease modelling and epidemiology, have indeed been made publicly available ({\em e.g.} \cite{centre_for_mathematical_modelling_of_infectious_diseases_covid-uk_2021, mrc_centre_for_global_infectious_disease_analysis_covidsim_2021, adam_kucharski_2020-cov-tracing_2021, adam_kucharski_2020-ncov_2021}). However, even these models still lack the transparent and readily traceable chain of evidence connecting data and assumptions with model outputs that would allow them, and their results, to be readily available and independently assessed. It is also commonly the case that data availability, coverage and quality are extremely variable or, if available, data may not be in a form that can be easily used without curation or transformation, further analysis, and a detailed description of the data. The ephemeral nature of some data sources and the rapid evolution of datasets used during an emergency, combined with the sparsity or absence of metadata describing datasets, all compound the problem of assessing evidence. In this paper, we examine the challenges around data coverage, quality and access with a particular focus on the issues and demands highlighted by the COVID-19 pandemic and outline the properties of a data pipeline designed to provide an infrastructure to address the demands of disease modelling for outbreak control policy-making, now and in the future.

Modelling work during the COVID-19 pandemic to generate estimates of key parameters and make predictions of likely outbreak trajectories has required multiple epidemiological models, operating at different scales and with varying levels of epidemiological detail (e.g \cite{kucharski_early_2020, giordano_modelling_2020, kucharski_effectiveness_2020, davies_association_2021}). This raises issues around the scale and resolution of the data used, and the extent to which the data are processed or abstracted prior to use. Moreover, the existence of multiple models, drawing on the same pool of available datasets, but in different ways, exposes a key point: that data, models and results are all research objects that require management.

From a practical perspective, the pandemic has made it clear that modellers must make the basis of their advice both transparent and accessible. Following the path from basic science to policy-friendly interpretation, via choice of parameter values, model structure, model assumptions, code implementation, and generation of outputs, is complex even for specialists. Version control tools like \texttt{git} combined with online repositories such as \textit{GitHub} have hugely enhanced sharing and collaboration on code, and managed repositories such as \textit{Zenodo} now exist for Open Science. However, this is just a small subset of what is required from a usable platform to support open and transparent epidemiological modelling, given the requirements that it be consistent in use across the range of likely applications, sufficiently unobtrusive that it is feasible for modellers to adopt it, and accommodating of the necessary diversity of data sources. Digital Research Infrastructure can now support transparent linking of the steps along this pathway.

We have developed a pipeline that provides an open and publicly accessible ‘chain of trust’ to transparently connect primary data to research outputs via open source, publicly accessible analysis and modelling code. This pipeline provides a route by which scientific hypotheses and study results combined with other sources of societal data (e.g. epidemiological, demographic, geographic and health service use) can contribute, through intermediate analyses, to publicly available and openly tested models while enabling generation of outputs whose dependencies can be fully interrogated.

In developing this pipeline, we provide an Open Science solution that addresses long-standing and critical problems in public health and livestock disease control. For example, Keeling \cite{keeling_models_2005}, in a review of the modelling effort during the UK Foot and Mouth disease outbreak of 2001, identified a number of sources of conflict associated with the use of mathematical modelling in emergency veterinary public health.  Keeling argued that tension between the veterinary and modelling sciences arose at least partially because of a disjunction of experience at different scales: that veterinary expertise was likely to be more accurate and effective at a local scale, whereas models were most effective in integrating the risks associated with multiple, larger scale events. It therefore becomes difficult to seek consensus across these groups, due to the confounding of perspective with professional expertise. A more general point was made by Matthews et al. \cite{matthews_neighbourhood_2003}, who argued that, as the spatial scale on which decisions and interventions are required increases, the threshold at which it is worthwhile to intervene, when measured in terms of the estimated risk of infection per premises, will tend to decrease. In this case also, we can note a problematic confounding of perspective (local, regional, national or supra-national) with the properties of any models operating at these different scales.  Unfortunately, it is all too easy to see how these systemic issues could translate into a lack of confidence in the results of a quantitative analysis, or even into a dismissal of modelling results where these conflict with the ‘common sense’ of an influential grouping.  

One way to overcome these potential problems is to seek to maximise the transparency of the modelling process, opening to general scrutiny the logic which has given rise to potentially contentious results, clearly reporting analytical assumptions and the provenance of the data resources which have underpinned the analysis.  To build scientific, political and public consensus, at a minimum, it is desirable to avoid mistakes arising from poor management or weak understanding of the data resources used. It is also important to avoid propagating any errors which do arise, and essential to have tools to find any such issues.  The importance of maintaining transparency and supporting better management of provenance of data, data products and modelling outputs is underlined by the official statement made by the UK Office for Statistics Regulation in November 2020 \cite{noauthor_osr_2020}.  Three key objectives were specified in respect of governmental use of data to support COVID-19 decision making: namely that \begin{enumerate*} \item where data are referenced publicly, that the data or at minimum the provenance of the data be published; \item where models are referenced publicly, that model outputs, methodologies and assumptions also be published; and \item where decisions are justified by reference to data, that this be made publicly available. \end{enumerate*}  These objectives are supported by the functionalities of the FAIR data pipeline described below.

This paper initially describes the issues we are trying to address and the existing tools that partially address them. It then follows the conceptual development and implementation of the open source pipeline to connect baseline assumptions and data to epidemiological models and their outputs. Using exemplar epidemiological models, we demonstrate how model runs are associated with a specific, cumulative chain of dependencies, supporting the critical examination of assumptions. If errors (or issues) are identified in primary datasets, analyses or modelling code, downstream model outputs can be automatically and transparently invalidated. The finalised infrastructure aims to provide an ecosystem for the epidemiological and wider scientific communities within which data, models and results can be managed in a transparent and publicly accessible way.

\subsection{FAIR research objects, provenance and data cataloguing}

The FAIR data principles\cite{wilkinson_fair_2016} were proposed as guidelines to apply to data, making them findable, accessible, interoperable and reusable so that both researchers and machines are able to find, access and (re)use data. The principles have been widely adopted and subsequently also been applied to other research objects such as software \cite{lamprecht_towards_2020}, workflows \cite{goble_fair_2020}, machine learning algorithms and executable notebooks; and guidance for ``FAIRification'' of data has been developed \cite{fairplus_fair_nodate, jacobsen_generic_2020}.

The main data descriptors required to achieve FAIR data according to the principles are: \begin{enumerate*} \item persistent identifiers, such as Digital Object Identifiers (DOIs, \cite{davidson_digital_1998}) for data, metadata and software, Open Researcher and Contributor IDs (ORCIDs \cite{noauthor_orcid_nodate}) for people, and \item standardised ways of recording metadata. \end{enumerate*}

As regards metadata standards, the open source pipeline requires a data registry that is aligned to existing formats and vocabularies. This will enable interoperability with other systems and support integrating and exchanging data in a straightforward way. To reduce the ambiguity in diverse data representations, we rely on common formats (such as JSON-LD \cite{noauthor_json-ld_nodate}), and common terminologies or FAIR vocabularies \cite{cox_ten_2021} that provide clear definitions and persistent identifiers for the terms. For example, we use  the \textit{provenance vocabulary} (PROV-O) to faithfully represent the entities, activities, and people involved in producing a research output.

\section{Requirements analysis}
\label{section:reqanalysis}
The Scottish COVID-19 Response Consortium (SCRC) \cite{scottish_covid-19_response_consortium_scottish_2020} was created as part of the Rapid Assistance in Modelling the Pandemic (RAMP) initiative, coordinated by the Royal Society \cite{royal_society_rapid_2020}. During 2020, the consortium comprised over 150 volunteers from multiple universities, research centres and industrial partners across the UK. During the design phase, we carried out a requirements analysis involving modellers, epidemiologists, data and policy experts, and software engineers from across the consortium to determine how such a pipeline would be used, document these use cases and extract from them the most important requirements. A search was made for existing technologies that might satisfy these requirements, but it yielded no results (see Section~\ref{section:related} for details), and a prototype pipeline was built to investigate the pros and cons of different approaches. In the light of this exercise, the current pipeline was then designed and built to meet the specifications.

\subsection{Use cases}
\label{subsec:usecases}

From the identified use cases for the FAIR data pipeline, ultimately 17 were taken forward and written up in detail. These use cases generated a variety of requirements, such as being able to query the pipeline to easily establish whether a dataset is already registered; being able to run analyses using the pipeline inside a Trusted Research Environment (TRE); being able to raise issues with data or software at run time or retrospectively to track the quality of these entities; and to inspect an output generated by a third party and identify whether any issues have ever been raised with any component in its provenance. An important criterion established in the use cases was that the software be easily approachable with low technical barriers of entry, thus making the pipeline accessible to end users.

\subsection{Requirements}
\label{subsec:requirements}

In the context of our objectives for a pipeline for epidemiological models, and mindful of the requirements specified (above) by the Office for Statistics Regulation, we require:

\begin{enumerate}
    \item a FAIR representation of the research objects involved, such as datasets and software, and the ability to trace updates to them, identify specific versions being used in analyses, track their provenance and integrate all the information necessary to understand how results are produced. This should include the ability to manually add provenance to the system where, for instance, a policy report contains figures, or simply results, generated by the pipeline;
    
    \item the ability to work seamlessly with data that is not publicly available, and indeed to be able to work fully offline, for instance in situations utilising sensitive data inside TREs such as  National Health Service (NHS) Safe Havens. In these situations it is still necessary to be able to uphold the FAIR principles and to allow for the evaluation of the provenance of research outputs by providing public access to metadata. However the data must be provably isolated from this process to comply with access requirements;
        
    \item interoperability with existing standards. No approach to solving these problems will be successful if it invents a series of new ``standards'', when a plethora of existing standards already exist. This is because so much data already exists in repositories that already comply with these standards, and users must be able to easily pull these datasets into the pipeline. This does not presuppose that existing standards have to be used to define internal formats, but at least there has to be the ability to interoperate through seamless import and/or export (e.g. DOIs, W3C-PROV \cite{groth_prov-overview_2013} and W3C-DCAT \cite{albertoni_w3c-dcat_nodate}); and
    
    \item that the system is not disruptive for end users, providing clearly identifiable short-term benefits to epidemiologists, and the Research Software Engineers working with them, to encourage uptake while causing as little friction as possible within their existing workflows.

\end{enumerate}

While reproducibility of results is desirable and may be possible where the data used are publicly available, it is not a core requirement since our concern with simplicity and user-friendliness (to minimise barriers of entry to modellers) can be in conflict with, for instance, containerisation approaches (containers package up code and all of its dependencies to allow quick and reliable transfer of analyses from one computing environment to another) that allow full reproducibility in all circumstances.

A number of other requirements were identified for individual use cases, but since they are mostly very specific to individual problems, and can be solved by (for instance) simply refining interfaces to help end-users use existing functionality, they are not listed here.

\section{Related work}
\label{section:related}

Our FAIR Data Pipeline must combine components for executing modelling workflows as well as recording the information and provenance of the research objects. This combination enables traceability of the modelling results.

In this section, we review related systems that provide similar functionality. We categorise them as \begin{enumerate*} \item those providing databases or online data repositories, often focused primarily on data, \item those providing ways of recording workflows, usually focused primarily on reproducibility, \item those recording the provenance of research objects by tracing code as it executes, and \item those providing or relying on version control systems, often combining software and data repositories. \end{enumerate*} Note however that there is significant overlap between these categories, so we have endeavoured to place the tools in the most appropriate category.

\subsection{Online data repositories and databases}
\label{subsec:data}

The ability to store data was not a core requirement of our data pipeline because of the range of storage solutions already available. Indeed, the specific storage mechanisms at individual sites, such as inside NHS Safe Havens, are well established and unlikely to change. Nonetheless, existing data storage solutions offer a partial solution to the issues that we are trying to address.

For instance, Zenodo \cite{european_organization_for_nuclear_research_zenodo_2013}, and other online repositories such as figshare \cite{figshare_figshare_2021}, provide persistent identifiers (DOIs) for any form of data, and also record standard associated metadata. By contrast, FAIRDOMHub \cite{wolstencroft_fairdomhub_2016}, and the underlying FAIRDOM-SEEK \cite{wolstencroft_seek_2015} software, provides a FAIR data and model management service specifically for Systems Biology, with metadata tailored specifically to this community.

Splitgraph \cite{noauthor_splitgraph_nodate}, on the other hand, is a PostgreSQL-based tool for building, versioning and querying reproducible datasets. It provides a public data store, while allowing provenance tracking of datasets that are created within Splitgraph. Dolt \cite{noauthor_dolt_2021} (together with the associated online repository, DoltHub \cite{noauthor_dolthub_nodate}) provides similar SQL functionality, with commercial options for private data storage, but does not generate provenance. Fully commercial data stores, such as data.world \cite{noauthor_dataworld_nodate}, also offer similar capabilities.

All of these data stores offer some of the functionality required by the pipeline, but all would require significant changes to users' workflows and none can be used offline. Nonetheless, the ability to interoperate with existing data stores like these, particularly public ones such as Zenodo, would be very valuable for accessing existing published datasets.

\subsection{Reproducible workflows}

As discussed in Section~\ref{subsec:requirements}, computational reproducibility was not seen as a critical requirement for the pipeline during the development of the use cases \cite{scottish_covid-19_response_consortium_data_2021}. However, a key requirement was the ability to trace exactly what code was run on which datasets to ensure accurate provenance recording. Traceability can be seen as an indirect element of a reproducible workflow, therefore there are overlaps with tools that already exist for these purposes.

Highly developed tools exist for managing and scheduling workflows \cite{freire_provenance_2008} that generally guarantee reproducibility, such as Galaxy \cite{afgan_galaxy_2018} for scientific workflows, Apache Airflow \cite{noauthor_apache_nodate} for more general data pipelines, and before that, Kepler \cite{noauthor_kepler_nodate}. Such tools are generally designed for complex and/or regular workflows, not the bespoke and one-off analyses that are more frequently produced in academia and which are detailed in our use cases.

On the other hand, Kaggle \cite{noauthor_kaggle_nodate} provides an online repository containing a wide variety of publicly available datasets and a cloud-based Jupyter Notebook \cite{noauthor_jupyter_nodate} environment for reproducibly analysing these datasets. However, the Jupyter-based workflow does strongly constrain what it is possible to do with the system, and the utility of cloud-based systems may also be undermined by legal requirements (e.g. General Data Protection Regulation (GDPR) \cite{information_commissioners_office_guide_2021}), which place requirements on data processors not to export data outside specific jurisdictions. Quilt \cite{noauthor_quilt_nodate} is an open source data hub that allows analyses to be run in (much more general) Docker containers \cite{noauthor_docker_nodate}. Quilt also provides a commercial product, QuiltData \cite{noauthor_quiltdata_nodate}, for managing private data. Neither of these explicitly allow provenance to be extracted, but the information is indirectly available through the contents of the notebooks and containers. A different issue with commercial platforms for data management and reproducibility is exemplified by FloydHub \cite{noauthor_floydhub_nodate}, which provided similar functionality to QuiltData, but the company that provided this service has ceased trading during the writing of this manuscript.

At least two tools were created during the pandemic specifically for reproducible analyses of COVID-19 data. Covid Model-Runner \cite{noauthor_covid_2020} was created in the early months of the pandemic to automate the epidemiological analysis of COVID-19 datasets that predicted future outcomes under different control scenarios. It used Docker to containerise the analyses and enforced a standard input and output schema for the models to ensure easy comparability. However, despite the excellent work that went into it, the tool was not widely adopted, perhaps due to its narrow focus, combined with the complexity of adapting code to use it. The lesson from this work may simply be that such tools must be adaptable to the workflows of the end user and cannot assume that the converse will apply, unless they provide a critical (to the epidemiologist) service not otherwise available. The second platform that has been developed during the pandemic has had a very different trajectory, providing as it does just such a critical service –– OpenSAFELY \cite{williamson_factors_2020, noauthor_opensafely_nodate} was developed to allow open and reproducible analysis of sensitive NHS patient data. It holds electronic health record data for 40\% of the population of England, and nine papers have already been published using the platform, providing important results about the disease and the efficacy of treatments. OpenSAFELY is designed for the analysis to be fully open, with all activity publicly logged. It can create its own TRE to operate appropriately on sensitive data or can be deployed on top of an existing TRE as a privacy-enhancing layer. It has many desirable features that enhance privacy such as the lack of access by users to raw data even when analyses are being conducted. However, it is expressly designed to operate solely inside TREs, and as such it puts strong constraints on how users interact with it, making it much less suitable for use on non-sensitive data.

These tools for reproducible workflow management provide useful functionality, but many require even more substantial changes to user workflows than the data management systems above. For specific tasks working with sensitive data, OpenSAFELY provides totally new levels of privacy protection, but outside that narrow focus, we believe that the constraints such tools impose are too heavy for our uses.

\subsection{Provenance Tools}

Data provenance, or pedigree or lineage, is about documenting the processes that produce the data in its current form. Much of the early work capturing provenance for scientific workflows is summarised in a conference demonstration \cite{davidson_provenance_2008} and a review \cite{freire_provenance_2008} in 2008. Since then, a standard has emerged for documenting data provenance -- the W3C Provenance Ontology \cite{groth_prov-overview_2013} -- and there are multiple tools that support recording provenance in different contexts. Here, as well as tools that capture provenance through workflow management (e.g. \cite{noauthor_apache_nodate, noauthor_kepler_nodate}, described above), we classify provenance tools into three further groups -- tools that work at the level of the operating system, tools that generate provenances for specific languages, and tools that manage provenance data for multiple languages.

Camflow \cite{pasquier_practical_2017} is an example of an operating-system-level tool, capturing ``whole system provenance'' \cite{pohly_hi-fi_2012}. It captures the relationships between Linux operating system kernel objects (such as files and threads) during execution and stores these such that they can be represented as a Directed Acyclic Graph (DAG). This is a very powerful tool, but captures so much information it is hard to use for a relatively ``simple'' modelling task. RDataTracker (and rdtLite and rdt) \cite{lerner_using_2018} are tools to collect provenance information from code executed in R, outputting to PROV JSON format. This can be visualised as a Data Derivation Graph (DDG) showing how data and computation led to an result. It requires no alteration to the underlying R code, providing a very low barrier of entry, but is inappropriate for our use both because it slows down the code significantly in order to trace it, and it only works on R code, strongly constraining the provenance that can be captured. recordr \cite{jones_recordr_nodate} is another tool working only in R to record provenance of analyses run in R without altering the source code. Unlike RDataTracker it does not significantly slow down the code, but is limited by only instrumenting a very limited number of function calls that read and write files. However, it works directly with the DataONE earth and environment online repository (\url{https://www.dataone.org/}), allowing metadata to be recorded for some of the data being used by scripts being traced. Finally, recipy \cite{jackson_recipy_2018}] is a provenance tracking tool for Python. Again it requires only minimal changes to the code to be tracked, but limited in both that it only tracks code in Python, and that it just traces file paths, and does not retain the data or store any metadata on the data being read and written. The Core Provenance Library (CPL) \cite{noauthor_core_nodate} provides and interface to a relational database that is used to store provenance information. It is written in C/C++ and provides interfaces in R, Python and Java as well. Developers integrate these interfaces into their code and actively use them to store provenance information. However, the provenance APIs provided are very low-level, not automatically capturing provenance as code is run, but rather providing the ability to manually create, look up and link objects in the database, making the barrier of entry just too high for most users.

These provenance tools are all very valuable in their own rights, and CPL is the closest of the tools to our requirements, but we believe that the barrier of entry for such a tool is just too high for our users, while the other, simpler approaches are too constrained in what they can track, and conversely the whole operating system approach just too complex for our needs.

\subsection{Version Control}

Source code for software has been curated through the use of version control software such as \texttt{git} \cite{noauthor_git_nodate} for many decades, and there is widespread adoption throughout the epidemiological modelling community (e.g. \cite{mrc_centre_for_global_infectious_disease_analysis_covidsim_2021, centre_for_mathematical_modelling_of_infectious_diseases_covid-uk_2021}). However, version control for data is less well developed, not least because, unlike source code, data often involves files that traditional version control cannot easily handle at present, whether due to limitations in memory, time, or disk space. To allow large and binary data to be more easily managed inside version control, additional functionality has been added to \texttt{git} through \texttt{git annex} \cite{noauthor_git-annex_nodate}, which uses special remote stores including cloud object storage to manage data outside the git repository. However, this is not trivial to use, and so other platforms have been built on top of \texttt{git annex} to allow end users to access this functionality more easily.

Some of the data management tools described in Section~\ref{subsec:data} provide forms of version control. These are often quite primitive in the sense that they do not identify specific changes to datasets, but only that the dataset has changed. Qri \cite{noauthor_qri_nodate} is an open source tool for data management, versioning and sharing. Versioning happens at the individual dataset level, with commits containing high-level metadata on the data as well as more detailed information on its structure. Pachyderm \cite{noauthor_pachyderm_nodate} is a platform for reproducible science with versioned data repositories. It too provides provenance at the level of data repositories that are used and produced by analyses, and reproducibility through Docker images and Kubernetes \cite{noauthor_kubernetes_2021} for container management. However, it is a commercial product that has job limits on the free community edition and it carries only limited metadata on the data it tracks. Data Version Control (DVC) \cite{noauthor_dvc_nodate} (built on \texttt{git annex}), and DAGSHub \cite{noauthor_dagshub_nodate}, which uses DVC for data management, \texttt{git} for software version control and MLflow \cite{noauthor_mlflow_nodate} for reproducibility, provides similar functionality to Pachyderm including provenance tracking, and is designed especially around machine learning workflows. DataLad \cite{halchenko_datalad_2021} is also built on top of \texttt{git annex} and provides similar functionality to DVC and Pachyderm, but has a broader scope.

We believe that the tools described above, especially the open source DVC/DAGSHub and DataLad platforms, come the closest of all of the identified existing products to satisfying our requirements. They provide provenance information (though it's not clear if manual additions can be made), they can handle private and public data, and they are relatively unintrusive in how users interact with them. However, they still fall short in several significant respects. Critically, it's not clear how to separate the metadata from the data itself, since \texttt{git} and \texttt{git annex} appear to manage both 'under the hood', and so it's not clear how to ensure public traceability once part of the provenance of a model output is marked as belonging to a private data store. The metadata they provide is also not clearly interoperable with formal standards; although the provenance information can almost certainly be standardised, metadata standards go far beyond that. Users and organisations interacting with the system should be traceable using persistent identifiers such as ORCIDs, and the data management system should track persistent identifiers for datasets such as DOIs (see Section~\ref{subsec:standards}, below, for further details).

All of the tools described above are valuable and some are widely adopted, with many obvious applications. Many positive lessons were identified from them in developing our pipeline, but they are all missing core features that make them unsuitable for our purposes.

\section{Overview of FAIR Data pipeline}
\label{section:overview}

\subsection{Standards and Interoperability}
\label{subsec:standards}

Our solution satisfies the requirements for both syntactic (referring to the formats used for the representation of the data) and semantic (referring to the vocabularies used for the representation of the data) interoperability of metadata by relying on the following standards and technologies:

\begin{description}
    \item[JSON-LD:] a standard format that extends JavaScript Object Notation (JSON) for Linking Data to allow automated navigation from one piece of Linked Data through embedded links to other pieces of Linked Data across the web; we use a JSON-LD representation for each of the entities, and for the provenance report  \cite{noauthor_json-ld_nodate};
    \item[W3C-PROV:] a vocabulary for the provision of information about people and activities, such as running code, that are involved in producing a data product in the pipeline, and representing the data provenance \cite{groth_prov-overview_2013};
    \item[W3C-DCAT:] a vocabulary for data cataloguing that enables us to describe datasets and facilitates the consumption and aggregation of metadata from multiple catalogues \cite{albertoni_w3c-dcat_nodate};
    \item[DOI, ORCID and ROR:] persistent identifiers for uniquely identifying digital resources, people and organisations (respectively), and providing associated metadata.
\end{description}

We will continue to work on expanding the metadata representation, provide export of data in the relevant formats, and plan to package the data products using the \textbf{RO-Crate} approach. RO-Crates \cite{sefton_ro-crate_2021}, or Research Object Crates, are a lightweight, JSON-LD-based approach to packaging research data with their metadata, providing a standard import and export format. 

\subsection{Flexible, easy to use and secure}

The pipeline must fit easily into existing workflows for data access, processing, modelling and analysis, and to support those likely to be required in a crisis situation, while ideally reducing, and certainly not adding to, the workload of users. Such workflows might involve exploratory work on a scientist's local computer, the need for code to run on HPC nodes without direct runtime internet access, or working within data safe havens and similar restricted environments.

Many of the platforms described above rely on cloud-based solutions or heavily constrain what workflows are possible, providing a barrier of entry sufficiently high that they will never be adopted by the target audience. Satisfying the requirements for both simplicity and working with sensitive data led instead to our adoption of a distributed architecture with local data registries that can operate fully autonomously on a laptop or in a secure environment, and which can wrap any existing workflow with minimal changes to ease adoption of the pipeline. Since these local registries contain only metadata, which is not disclosive, they are able to synchronise with remote registries to satisfy requirements for public accessibility of provenance information even for sensitive analyses.

\subsection{Trust and quality}
During the RAMP period, SCRC proposed a model evaluation framework for Open Epidemiology that would ensure information about provenance, quality and robustness of modelling results would be available alongside any advice or reports that may be used in decision making. This would cover the key elements contributing to the validity of modelling results including the quality of the underlying science, confidence in the correctness of the software implementation and the reproducibility of outputs from it, the existence and results from validation of models and inference procedures, and the quality and provenance of data, combining it into an overall assessment of output policy-readiness.

The FAIR data pipeline enables such evaluation reports to be attached to objects in the provenance chain. SCRC's lead Research Software Engineers developed a software checklist \cite{scottish_covid-19_response_consortium_software-checklist_2020} intended to be completed by software developer(s) and updated for each release of their software. It can then be associated with the release through a dedicated table in the data registry. This checklist has been completed for key components of the FAIR data pipeline with copies stored as \texttt{software-checklist.md} in the root of each GitHub repository alongside the software source code.

\subsection{Overview of components}
\label{subsec:components}

\begin{figure}[ht]
    \centering
    \includegraphics[scale=0.5]{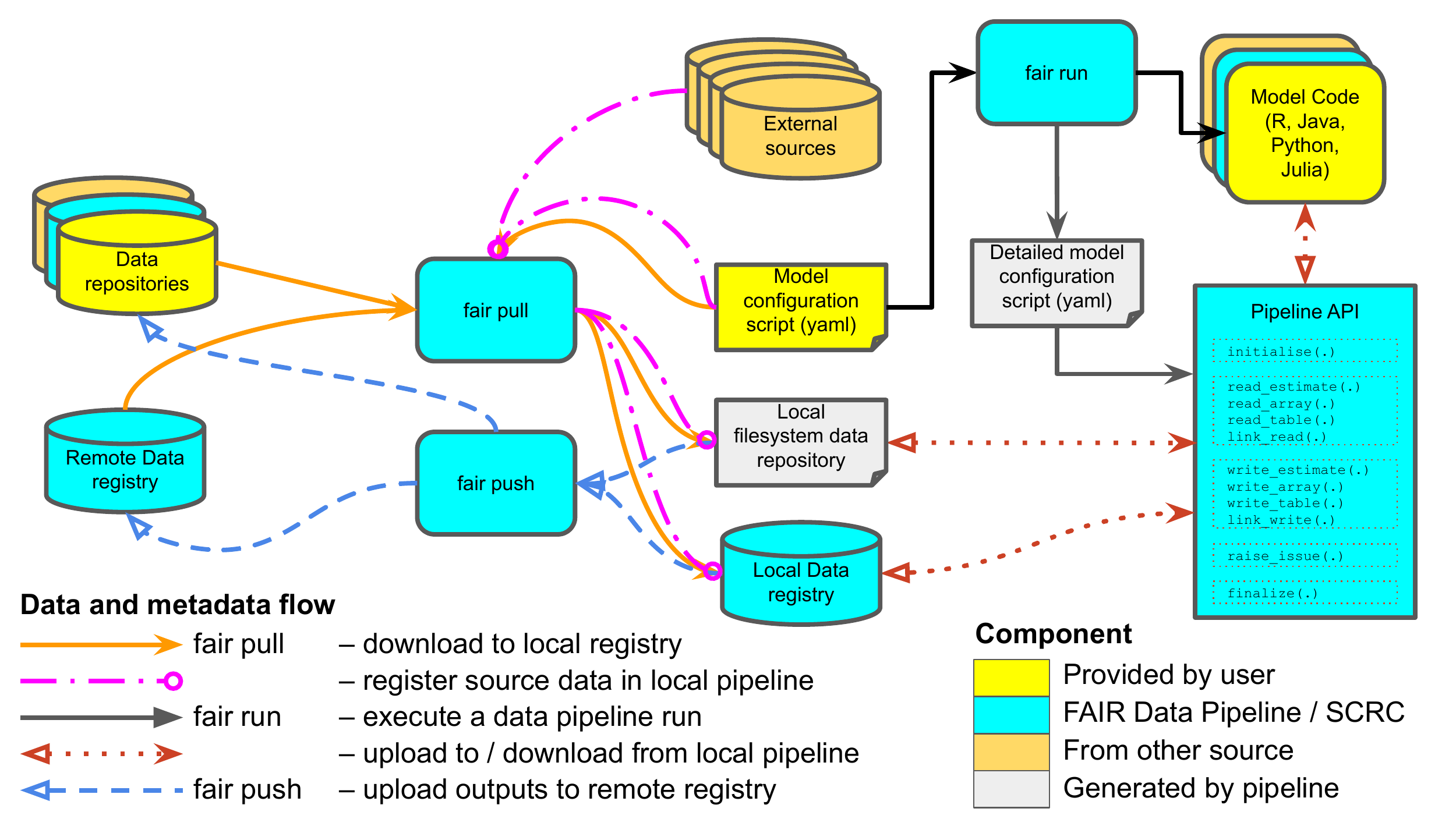}
    \includegraphics[scale=0.45]{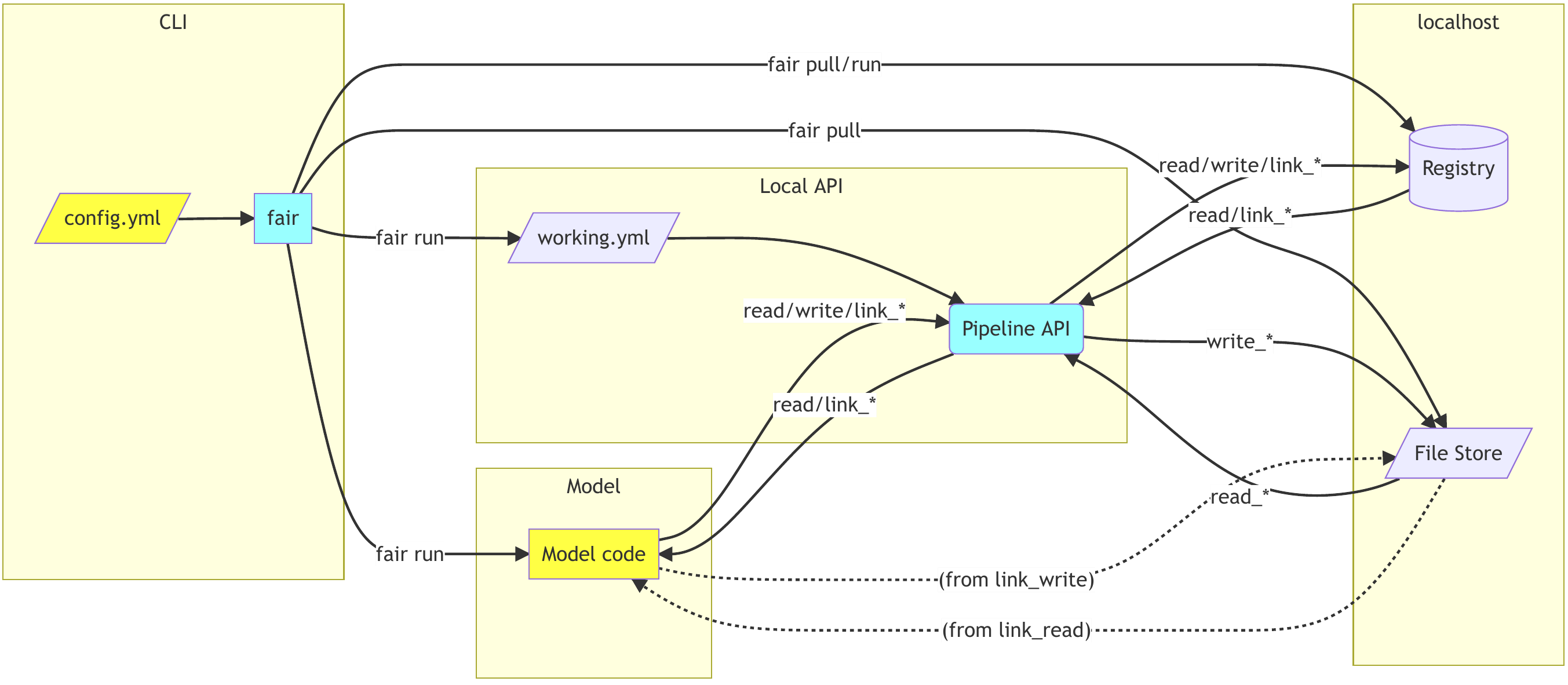}

    \caption{\textbf{Design of the FAIR Data Pipeline}. Upper panel: Control flow for whole pipeline showing local and remote registries, and actions that can be taken to populate and synchronise them.
Lower panel: Control flow for local pipeline, showing actions that can occur during a model run.
}
\label{fig:pipeline}
\end{figure}

The FAIR Data Pipeline software suite (Table~\ref{table:softwarelist}) consists of \begin{enumerate*} \item the \textbf{data registry}, a Django database-backed web application holding metadata and providing REST APIs; \item \texttt{fair}, a command line tool that can both synchronise metadata (and optionally data) between the execution platform and a remote data registry and is used to start experimental runs directly; and \item a set of language-native programming interface packages which can be added as dependencies of modelling software to enable it to read and write data from the pipeline. \end{enumerate*} See Table~\ref{table:softwarelist} for package details, and Figure~\ref{fig:pipeline} for how metadata and data flows through the pipeline. Provenance is tracked automatically by launching analyses through a command-line tool, and tracking files as they are read and written by minimal editing of the modelling code to wrap the read and write calls.

\begin{table}[!ht]
\caption{FAIR Data Pipeline Software: all packages are available under open source licenses and are developed as public repositories within the FAIRDataPipeline GitHub organisation \cite{scottish_covid-19_response_consortium_fairdatapipeline_2021}, and where appropriate, released through language-specific package registries. $^*$Package to be added.}
\label{table:softwarelist}
\begin{tabular}{lllll}
\hline
\textbf{Repository Name} & \textbf{Language} & \textbf{Registry} & \textbf{Package Name}\\
\hline
data-registry \cite{blackwell_fair_2021}                & Python (Django)  & N/A    & N/A \\
FAIR-CLI \cite{zarebski_fair_2021}                      & Python           & PyPI   & fair-cli \\
rDataPipeline \cite{mitchell_rdatapipeline_2021}        & R                & CRAN$^*$  & rDataPipeline \\
DataPipeline.jl \cite{mitchell_datapipelinejl_2021}     & Julia            & General& DataPipeline \\
javaDataPipeline \cite{boskamp_javadatapipeline_2021}   & Java             & Maven  & org.fairdatapipeline \\
pyDataPipeline \cite{field_pydatapipeline_2021}         & Python           & PyPI  & data-pipeline-api \\
cppDataPipeline \cite{field_c_2022}       & C++              &   N/A  & libfdpapi \\
\hline
\end{tabular}
\end{table}

\section{FAIR data registry}

The \textit{data registry} is a dynamically updated database that stores metadata associated with the diverse types of data utilised and generated within a typical epidemiological modelling workflow. In this context, we refer to data in the broad sense as any files, including scripts, code, figures and individual parameters. The system design anticipates a variety of different research objects being present in the modelling workflow, whose interactions give rise to new objects which are automatically entered into the registry. The registry was designed to hold the following research objects:

\begin{description}
    \item[datasets] either sourced from a data provider, such as a government department or agency, as open data accessed from another researcher, or generated as part of a registered workflow;
    \item[version-controlled processing scripts] typically written by the team using the registry, to 
    \begin{itemize}
        \item restructure datasets, including activities such as identifying anomalies and raising issues, data cleaning, selecting subsets, and linking records across datasets, 
        \item carry out curation activities on datasets,
        \item carry out specific analysis, producing output that is either going to form part of another output in its own right, or numerical quantities (typically a vector of parameters) which will be used downstream in the workflow;
    \end{itemize}
    \item[version-controlled analysis scripts] typically written by registry users;
    \item[analysis outputs] generated by the application of an analysis script to a dataset;
    \item[model parameters] either extracted from an analysis output, or entered into the registry as a quantity derived from the work of the wider scientific community, such as a parameter cited in a paper or a report from another modelling group;
    \item[version-controlled modelling code] typically written by registry users for use in generating specific model outputs;
    \item[mathematical model outputs] generated by use of a version of a mathematical model codebase, probably making use of multiple datasets and model parameters;
    \item[reports] generated by pipelining model and analysis outputs into a pre-specified report format, or generated manually (e.g. in Word) using such outputs.
\end{description}

The metadata associated with these research objects can be either intrinsic or extrinsic.  Intrinsic metadata includes fields which contain information relating to provenance. Data objects that enter the registry from outside the data pipeline have information uploaded to detail the source of the material, preferably including a persistent identifier such as a DOI as a commonly recognised, persistent method for machine actionable, globally unique identification.  Data objects that enter the registry having been produced by researchers working within the data pipeline will have metadata associated with them automatically detailing author and versioning. Such outputs will also trace their history through pipeline interactions to uniquely identify the provenance of the new objects. The operation of the data pipeline will not lead to any revision of the intrinsic metadata associated with a data object; these may need to be updated if further information about provenance becomes available, but this reflects a change in the user’s knowledge, not in the actual nature of the data.  A key property of the registry is that changes in the metadata associated with a data object will propagate to the provenance metadata associated with offspring data objects.  This is useful in maintaining consistently valid, high-quality intrinsic metadata across the entire population of research objects in the registry, but it is even more important when considering extrinsic metadata.

By contrast, extrinsic metadata can be updated over time.  The data registry includes two key elements of extrinsic metadata: one (``QualityControlled'') is an assessment of the quality, or fitness-for-purpose, of the research object, whether data or code. The other (``Issue'') can be used specifically for raising concerns about problems identified in the data or code, either at the point of data upload or generation or through later analysis. The status of the dataset is dynamically propagated through the registry, and hence is visible in the provenance of outputs generated using these datasets.  In a similar way, datasets that have simply been superseded can also be flagged, and reports and other outputs based on them can then be identified.  If such a report is used as the basis of (say) a briefing to government policy makers, it will be important to make it clear where it is indeed based on outdated information.  The key outcome is that all the information required to contextualise the work is available in the provenance metadata.  It is instructive to compare this situation with the more extreme case where a dataset is (say) discovered to have contained erroneous data for a period of time.  All versions of the dataset subsequently identified as erroneous can be flagged as invalid.  All outputs derived from these datasets will have this invalidation in their provenance; reports can be withdrawn, and there should be no risk of any work subsequently using these invalid data without this being apparent in the provenance metadata. In this way, the registry is the primary source of information to derive retrospective provenance, i.e. a detailed log of the execution of the computational task, including user-defined provenance in the form of annotations \cite{freire_provenance_2008}. 

The full schema of the registry database, including other metadata such as information about software releases and DOIs for published data, is available in Supplementary Figure 1.

\section{Examples}

We have selected three examples to demonstrate features of the pipeline. These are not intended to push the limits of the software framework, but to provide simple to understand examples that can be replicated by the reader to get a feel for the complexity of using the FAIR Data Pipeline. The first is the reproduction in R of a simple Susceptible-Exposed-Infected-Recovered-Susceptible (SEIRS) epidemiological model used by Bjornstat et al. \cite{bjornstad_seirs_2020} to demonstrate disease dynamics. In the second example we reimplement this model in all four native languages of the data pipeline (R, Java, Julia and Python) and cross-validate the results. Finally, we show a more complex, but nonetheless very simplified, time-varying model of COVID-19 dynamics with parameters extracted from English epidemiological data from the pandemic, and pull these into the pipeline and run a deterministic simulation of the pandemic using those inferred parameters.

\subsection{SEIRS model}

\begin{figure}[ht]
    \centering
    \includegraphics[scale=0.7]{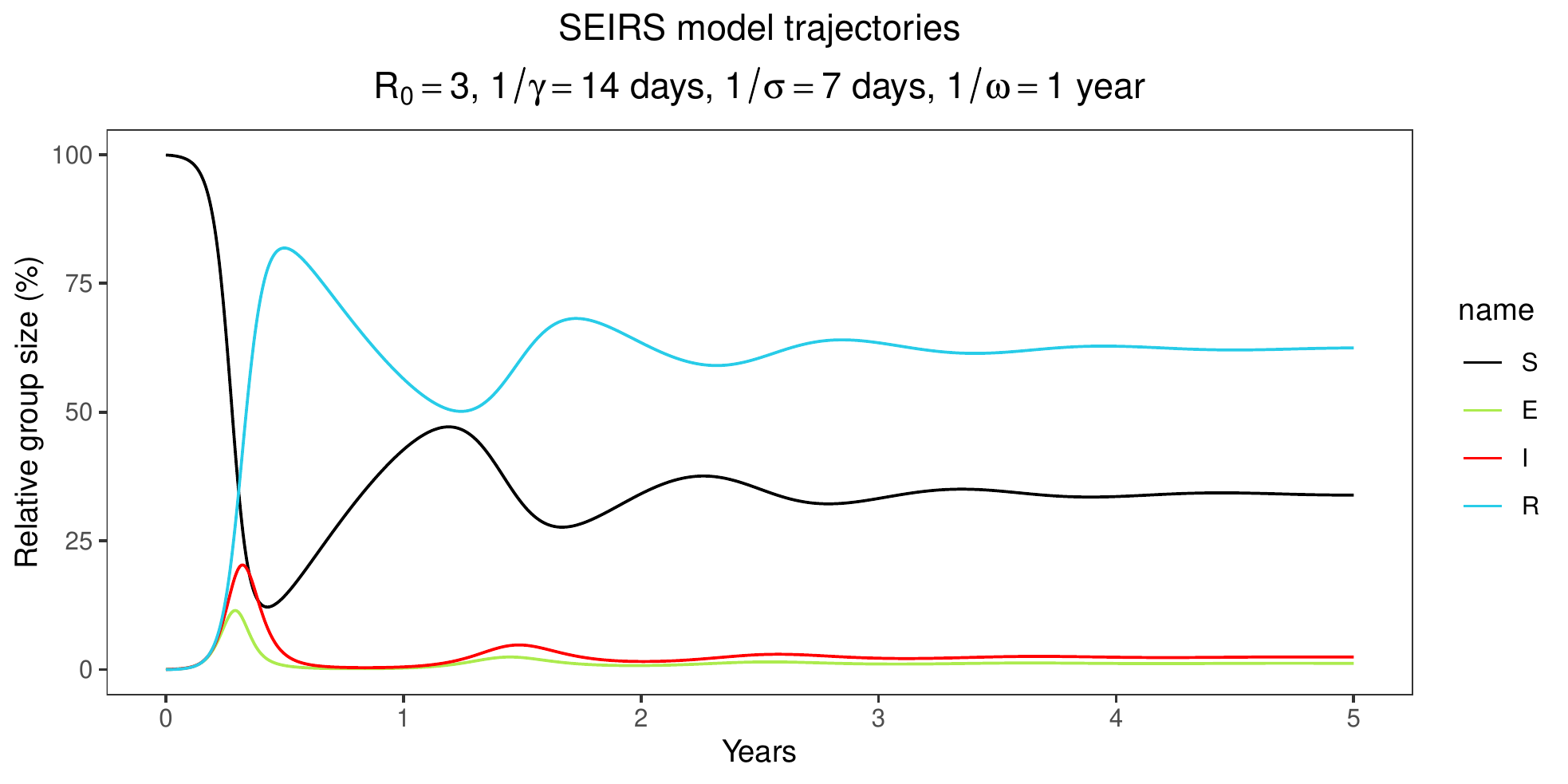}
    \includegraphics[scale=0.65]{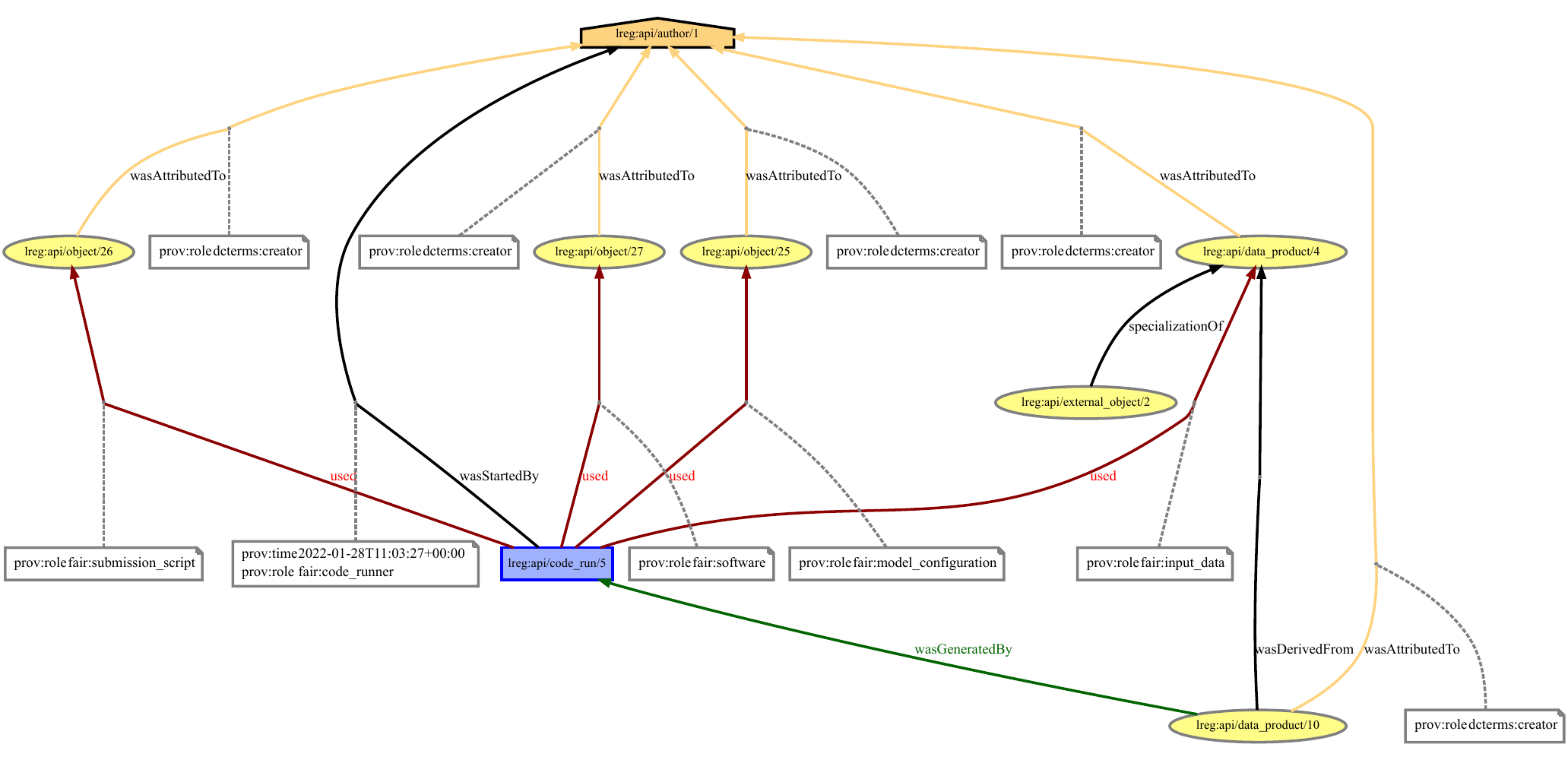}

    \caption{\textbf{Running an SEIRS model in R through the data pipeline}. Upper panel: Output of the model, showing SEIRS dynamics matching Bjorstadt et al. \cite{bjornstad_seirs_2020}.
Lower panel: Provenance of model output, tracing SEIRS plot back to parameter inputs. Note that provenances are ordinarily provided in standard PROV-O format, hyperlinked to the research objects being traced and with additional metadata, but they are simplified here for display purposes.
}
\label{fig:simpleModel}
\end{figure}

To demonstrate a simple epidemiological model being run through the pipeline, we take the example of the SEIRS model used by Bjornstat et al. \cite{bjornstad_seirs_2020} and reproduce the results that lead to Figure 1b in that paper. The full code is available online in an R package \cite{mitchell_rsimplemodel_2021}, which provides instructions for how to run it in the \texttt{README.md} file on GitHub. A vignette is also provided \cite{mitchell_seirs_2021} showing the fully worked example with results. \texttt{fair pull} is used (see Figure~\ref{fig:pipeline}) to populate the local registry with the parameters from the source manuscript (using a \texttt{register} block in the configuration file to ensure the data is in the pipeline). \texttt{fair run} is then used to execute the R script (the R code is executed from the root of the git repository via the \texttt{script} block of the configuration file). The R code itself is only very lightly edited from an equivalent non-pipeline version, with the addition of \texttt{initialise()} and \texttt{finalise()} steps to start and stop the pipeline monitoring, and the replacement of hard-coded file paths with calls to \texttt{link\_read()} and \texttt{link\_write()} with references to labels given in the configuration file (by default the unique names under which they are stored in the registry). These functions simply return appropriate paths, so can be directly used in place of filenames in any normal R code. The output of this simulation is shown in Figure~\ref{fig:simpleModel} along with the provenance of the output figure.

\subsection{Model comparison}

\begin{figure}[ht]
    \centering
    \includegraphics[scale=0.7]{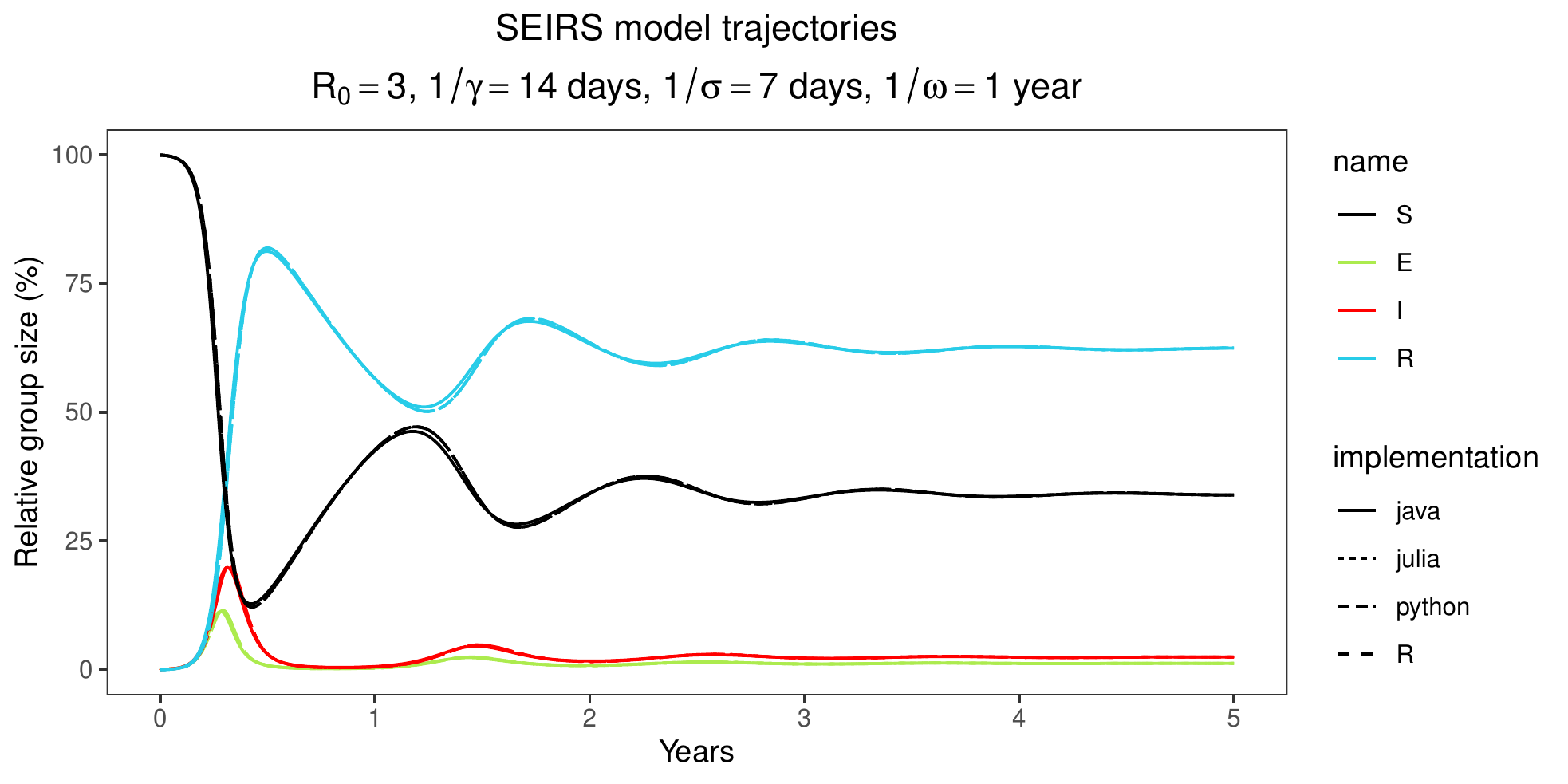}

    \caption{\textbf{A comparison all four language implementations through the data pipeline.} The figure shows the plots as almost superimposed. The reason for the slight discrepancy was automatically identified during the cross-validation, and in consequence, the provenance of the figure highlights the issue raised.}
\label{fig:comparisonModel}
\end{figure}

The SEIRS model described above was implemented in all four of the native languages of the data pipeline, and the Java \cite{boskamp_javasimplemodel_2021}, Python \cite{field_pydatapipeline_2021} and Julia \cite{mitchell_datapipelinejl_2021} implementations can be found online. We ran all four models through the pipeline together and then wrote a small cross-validation script to compare the models. The comparison is shown in Figure~\ref{fig:comparisonModel}. Critically, the models disagree due to a difference in timesteps used and the length of a year in the implementations (365 days vs 365.25 days). Examining the provenance of the figure shows that this was automatically identified by the cross-validation code, which then added an issue to the Java model output (the issue can be seen in the registry interface in Supplementary Figure 2). This issue can then be traced through the provenance to anything that uses this data product. This ability to trace problems with downstream research objects (in this case Figure~\ref{fig:comparisonModel}) resulting from (even retrospectively) identified problems with upstream data is a core strength of the pipeline.

\subsection{COVID-19 model}

\begin{figure}[ht]
    \centering
    \includegraphics[scale=0.4]{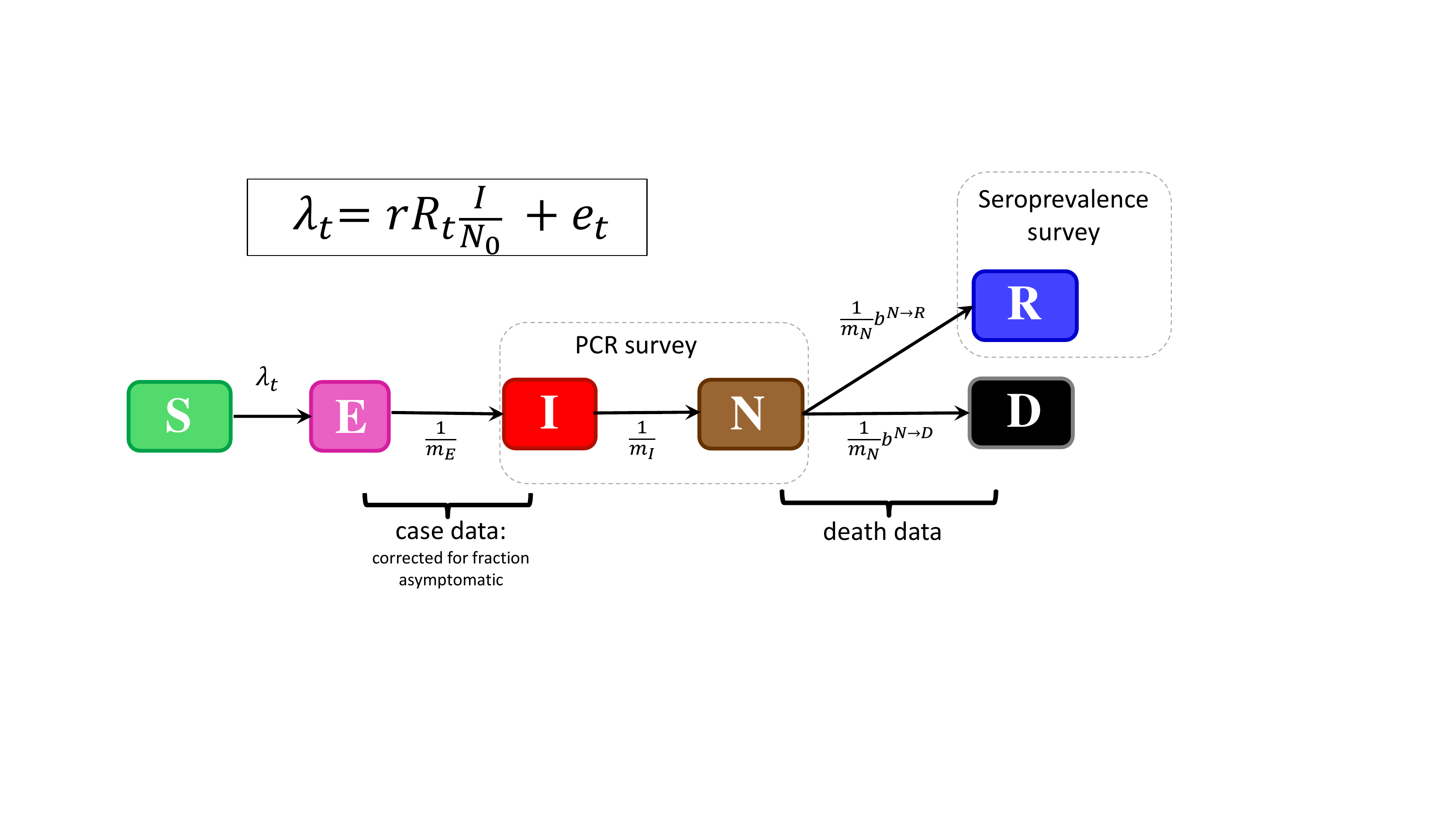}
    \includegraphics[width=13cm]{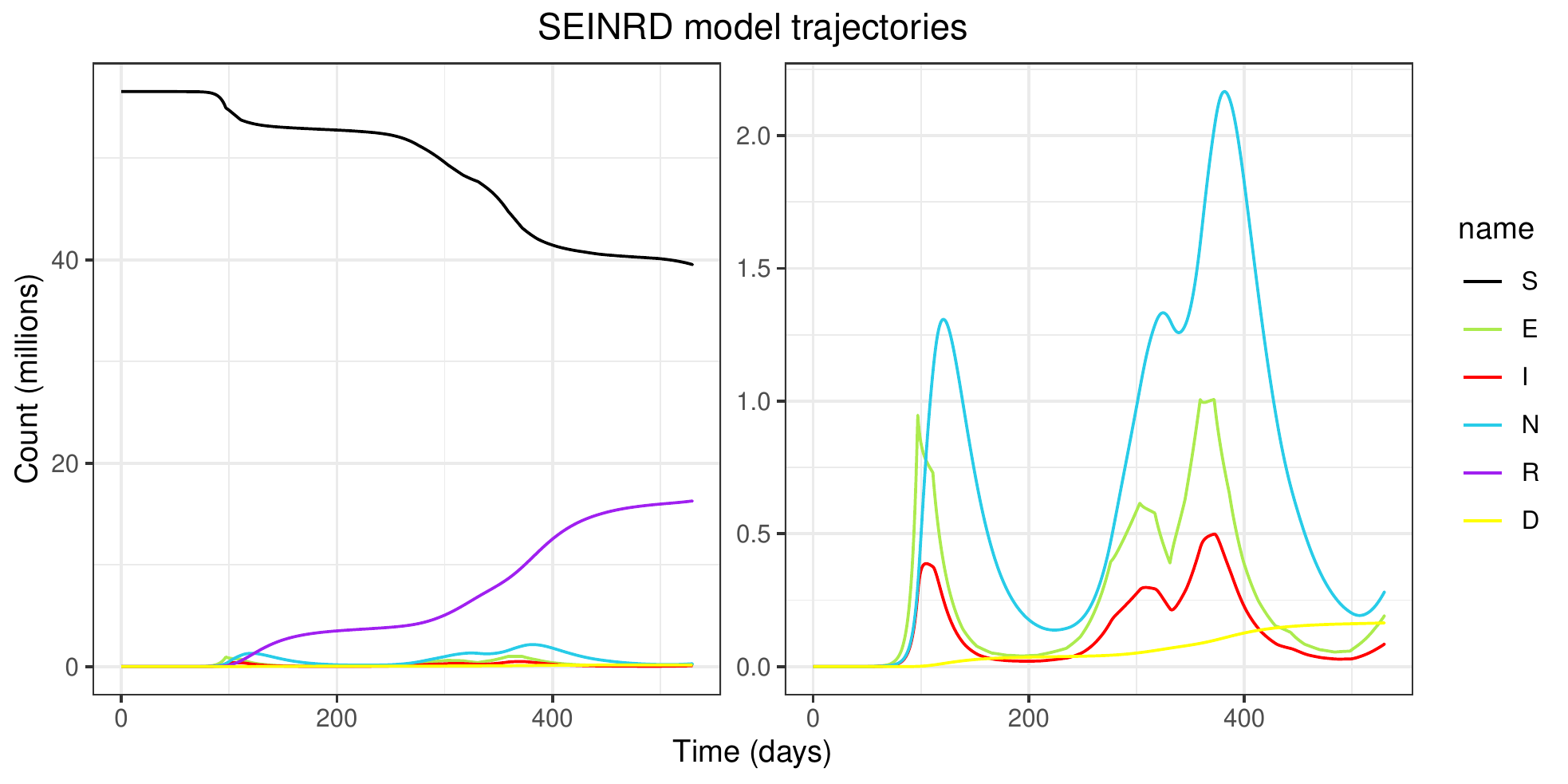} 

    \caption{\textbf{Epidemiological model fitted to 2020-21 COVID-19 data from England}. Upper panel: This shows a homogeneous model (i.e. with no age or spatial structure) with compartments representing, for each point in time, the number of individuals who are susceptible $S$, infected with COVID-19 but not infectious $E$, infectious $I$, isolating/non-infectious $N$, recovered $R$, and those who have died $D$. The figure shows the per-capita rates of transition for each allowed transition between states. The parameters $m_E$, $m_I$. $m_N$ represent the average time spent in the states $E$, $I$ and $N$ respectively, whilst $b^{N\rightarrow D}$ and $b^{N\rightarrow R}=1-b^{N\rightarrow D}$ are the probabilities of death and recovery. The force of infection $\lambda_t$ is dependent on: a per-capita external force of infection $e_t$; and, a frequency dependent term $r R_t I /N_0$ that represents the rate of disease transmission in terms of the average time spent in an infectious state (here $r=1/m_I$) and the real-time reproduction rate $R_t$. Both $R_t$ and $e_t$ vary with time and are driving variables for the model. As indicated, case and death rate data inform the transitions shown whereas PCR and seroprevalence data from the Coronavirus Infection Survey \cite{offices_of_the_nuffield_professor_of_medicine_covid-19_2021} informs on the numbers of individuals in the compartments $I+N$ and $R$ respectively. With the exceptions of $m_E=4$ days and $m_I=4$ days, all model parameters were determined via Bayesian inference using the methodology described in \cite{pooley_estimation_2021} applied to a stochastic version of the model described here. Note that, although the latent and infectious periods are fixed, changing these quantities has the effect of rescaling the inferred $R_t$ about 1, but does not impact on the other parameters significantly. Lower panel: The deterministic SEINRD R code captures the first, second and third wave of the outbreak in terms of cases and deaths attributed to COVID-19.}
\label{fig:complexModel}
\end{figure}

Epidemiological models such as the one described above are critically dependent on the values of parameters; these are typically difficult to estimate and subject to variable levels of uncertainty. It is therefore important to be able to trace model outputs back to the parameterisation(s) used to generate them. We illustrate this point in a case study in which the parameterisation  was determined by fitting the epidemiological model shown in Figure~\ref{fig:complexModel} to COVID-19 epidemic data from England up to mid-2021. These include the static parameters (see Figure~\ref{fig:complexModel} (upper panel)) and the time varying external force of infection $e_t$ and basic reproduction number $R_t$ that account for the impact of pandemic response (e.g. lockdowns and travel restrictions) on the outbreak dynamics. The model runs shown in the lower panel of Figure~\ref{fig:complexModel} are generated by using the FAIR data pipeline to link the deterministic,  R-based implementation of the model to the parameter values generated via Bayesian inference (see Figure~\ref{fig:complexModel} caption). The full code is provided in the same git repository as the first example \cite{mitchell_rsimplemodel_2021}, and a vignette is also provided showing a fully worked example of it being run in the pipeline \cite{mitchell_seinrd_2021}. Where the Bayesian inference is itself carried out within the data pipeline, the provenance information will itself chain back to incorporate both the stochastic model and the primary datasets.

\section{Discussion}

During the COVID-19 pandemic, media outlets have channelled highly charged and politically polarised arguments about the trustworthiness of scientific advice for government policy and also of the advisors themselves, as well as debating the extent to which governments are, in any event, following such advice. While some such controversies are inevitable, as a scientific community endeavouring to provide the best advice we can to policy makers, we are at a disadvantage if the detail of our results is hidden and if the evidence chain that connects our advice to the data and models that underpin it is not just unavailable to the public, but in fact largely nonexistent. This situation has arisen although standards have been available for many years, promoting openness and reuse of data and metadata, in particular through the FAIR principles for scientific data management \cite{wilkinson_fair_2016}.

The FAIR Data Pipeline was SCRC's response to this aspect of pandemic response. Querying epidemiologists involved in both human and animal disease modelling, we could identify no tools being used that satisfied either FAIR principles or which publicly presented the provenance of research outputs. We felt that the media storm surrounding some of the scientific advice during the pandemic demonstrated that such a tool would be valuable in improving trust in science used for public policy. In addition to the indirect and intangible costs arising from of diminished public faith in science, there is also potential for wider efficiency benefits to accrue from developing and using software to support FAIR data management.  An analysis of the qualitative and quantitative costs of not having FAIR data has estimated a negative impact of \euro{10.2bn} on the European economy \cite{european_commission_directorate_general_for_research_and_innovation_cost-benefit_2018}.

Accordingly, a requirements analysis was run to determine what capabilities were needed from such a tool -- specifically, what functionality a data pipeline would need to provide to be useful to not just to the epidemiological and other modelling communities, but also the broader lay and policy audience \cite{scottish_covid-19_response_consortium_data_2021}. After a review of software available for data management and reproducible research, we concluded that no suitable tools existed in the public domain that could be easily adapted to these uses. We therefore developed the new open source suite of tools for FAIR management of data and models described here to improve the openness of science for policy and better support the traceability of evidence, with as low a barrier of entry as we could devise. Although a small amount of friction remains, our intention has been to ensure that there are benefits for all potential users that can be realised in only a few minutes: for instance for modellers, the ability to automatically trace the provenance of the output of an existing model with only a single short configuration file to describe any input and output data, and minimal changes to the code; for a data manager or user to look up a file in a data registry and see what (if any) metadata is held on it, and what other versions of that data product also exist; or for third parties, the ability to look up a model output in the same data registry and see its provenance immediately, linked directly back to the code and data used to create it.

Critically, these tools are as non-prescriptive as possible in terms of how, in what environments, and with what software users carry out their analyses and modelling, while nonetheless tracking in fine detail exactly what versions of what data are ingested, what commits of what software are run and by whom, and what research outputs are generated. As well as automatically generating detailed provenance information, the pipeline also annotates the runs with other detailed metadata, and relies on existing standards to maximise the FAIRness of the data produced and to enable interoperability with other resources. Finally, where possible, it reduces the burden on users of correctly annotating data provided from external data providers (or even their own files) by retrieving metadata (and even the data itself) directly from the sources. Where this is not possible, it provides a simple text-based format to manually upload the necessary information in a straightforward manner.

Isolating metadata completely from the data being analysed and used means that this resource (in the form of data registries that can themselves be run directly on user laptops or created in the cloud and shared across research groups if desired) can be made publicly available even when the underlying data is highly sensitive, as usually the metadata and data have different licenses. When allowed by the metadata license, we can include it in the registry since the data is only identified through the checksums of the files (to ensure traceability) and any other metadata that the user chooses to upload.

By using standard formats and vocabularies, we are maximising the FAIRness of the metadata stored in the registry, and enabling interoperability with other resources. This also allows us to export the metadata in recognised and proven formats with existing tools to manipulate them, and the process of aggregating data should be simplified. However, there are still many challenges when including external resources and aggregating the metadata. Where the data is openly available, some sources provide data (e.g. CSV files) without any description of what the data means. Other sources provide data dictionaries, and this improves our interpretative ability when mapping the data into our registry, but, nevertheless, the mapping process is manual, error-prone and time consuming. By promoting a more standardised approach, we hope that data providers in epidemiological modelling and other domains may follow our lead in adopting the standards for data description that we support, which are widely used in other application areas.

\section{Conclusions}

\subsection{Trust, but verify}

By enabling the public release of the provenance of scientific policy advice, we believe that this data pipeline will allow users to be open about their work in a way that can only increase trust in well founded scientific conclusions. It will increase trust by allowing verification to take place more easily, and it will allow users to more easily identify potential problems in the logical constructs that have led to their conclusions through the integration of the issue tracker into the provenance system. With climate change a ongoing crisis, subject to sectional political and wider societal argument, and with critical inputs to any solution needing to come from the scientific community, the COVID-19 pandemic will not be the last time that we will be challenged on our openness and trustworthiness.

A situation where a domain expert delivers policy-relevant model-derived evidence, either without all of the choices made in generating this evidence being made explicit, or without providing supporting evidence for these choices, is clearly problematic in terms of public trust.  In practice, in the short term, there may be insufficient time to deliver either comprehensively.  Alternatively, where decisions have to be made rapidly, potentially based on uncertain and rapidly changing information, a key functionality is to ensure that choices are documented, to facilitate ongoing assessment of their validity, by the modellers themselves as part of their own scientific processes, but also by a wider population of scientific peers, and to a more limited but still valuable extent, by science-policy brokers, policy makers and the wider general public.  Although the latter will not necessarily have the time or technical background to assess the technical validity of modelling and analysis assumptions, trust can nevertheless indirectly be promoted by facilitating technical scientific challenge by those who are most equipped to do so.

It is also desirable that, as part of the provision of expert advice, domain experts are facilitated in curating a record of the choices they have made, and of their evolving understanding of these.  In particular, ongoing assessment of the validity and relevance of model releases and key data resources should be facilitated, whether by the modeller or by informed stakeholders.  In particular, where advice changes in the light of new information, support for resulting shifts in policy will be enhanced by an ability to demonstrate the link between new assumptions and new conclusions.  Other, more immediately pragmatic requirements include the ability to identify analyses and outputs which depend on outdated code or data; an ability to validate or invalidate a past analysis conditional on the current status of its underlying assumptions and data; and an ability to retrospectively reconstruct the validity of an analysis at a previous point in time, given the status of the underlying assumptions at that time.  We anticipate that these key functions will be particularly important to those specialists brokering evidence across the science-policy interface. Where the domain expert makes metadata detailing the provenance of a dataset or codebase available, the interpretation of these in terms of applicability and validity can be assessed and challenged by their technically proficient peers.  So long as the domain expert is updating the data registry to reflect changes in their own perception of the validity of assumptions and data resources, any other user can use the data pipeline to meet the operational needs described above or simply to explore the links between outputs and assumptions over time.  Thus, in general, use of the data pipeline facilitates good curation of metadata by the modeller, supports informed evaluation of work by scientific peers, and democratises access by the wider community to information about model assumptions. In so doing, we believe that a tool such as the data pipeline can help maintain public confidence in scientists and scientific work at the high level which best supports society and its needs.

\subsection{Future work}

\subsubsection{Documentation and Validation}

The plan to operationalise the pipeline is to integrate a suite of realistic policy-oriented models into the data pipeline.  The use cases previously described in Section 2(a), detailing a wide range of activities likely to be carried out by identified users (including mathematical modellers, science-policy brokers, policy-makers and the wider public), will each be implemented for the integrated mathematical models, as part of a process of analysing user-software interactions and developing documented procedures. We would hope to involve science-policy brokers in this process; their involvement will be invaluable, in that they are a key target user group, and can also plausibly serve as proxies for policy-makers and the general public in the process.  In particular, use cases 9-13 are different inspections of data and results that they (and other individuals like members of the public) might wish to make to understand the origins of conclusions that researchers present. Currently the data registry's web interface to address these use cases ({\em e.g.} Supplementary Figure 2) is limited, but further work is underway to improve this. Tools for provenance visualisation are also limited, and we believe further work is needed in this area to reduce the complexity of the diagrams produced, and increase their ease of use for exploration of data and results. If gaps become evident in the portfolio of use cases, these will be documented and carried forward for further attention. In the longer term we intend to pilot uptake in groups delivering model-based evidence to policy; it is likely that initial implementation and evaluation would best be carried out as part of an emergency simulation exercise, where the utility, costs and robustness of the data pipeline could be assessed within the context of the wider demands made of the scientists by policy-makers.

\subsubsection{Metadata Automation}

At the moment, several aspects of pipeline use are manual where they could be automated. These include \begin{enumerate*} \item the attribution of authors to software, which could be taken from \texttt{CITATION.cff} or \texttt{.zenodo.json} files, or other metadata in the git repository; \item the integration of metadata (including authorship) into the pipeline from DOIs associated with data as it is ingested; \item the automatic creation of Issues and CodeRepoReleases in the pipeline from issues raised and releases created in GitHub; \item the automatic generation of persistent identifiers (e.g. DOIs) when ExternalObjects are created; and \item improving the ease of syncing metadata between local and remote registries. \end{enumerate*} Integrating this functionality into the pipeline is desitable, since it will further increase ease of use, reducing the barrier of entry for new users.

\subsubsection{Interoperability}

Further work is also necessary on interoperability, to increase the FAIRness of the data managed by the pipeline, in particular to make it easier to catalogue, search, access and reuse metadata. The pipeline can already export metadata in PROV-O and JSON-LD formats for provenance and linked data, with some descriptions of the data products using the DCAT vocabulary. These representations can be extended to include more details on the different research objects as required. In addition, we also need to be be able to export the whole registry in a DCAT-compliant way for interoperability with other data catalogues, as well as exporting whole research objects with their provenance and other metadata in RO-Crate format. This should also allow us to import data with associated metadata directly from other platforms, either directly or via specifically created mappings to recognised standards. Finally, we need to integrate the ability to use different storage engines as backing stores for the pipeline.

\section*{Data Access}

All software is openly available on GitHub, archived on Zenodo, and (where appropriate) registered on language-specific package registries. See Table~\ref{table:softwarelist} for further details.

\section*{Author Contributions}

Contributions were determined using the CRediT taxonomy -- \url{https://casrai.org/credit/}.
Conceptualization: AB, ANGB, AR, DM, GM, IJM, LM, RD, RR, RT.
Data curation: AL, ANGB, CM, IJM, JE, JH, JW, MS, NC, PB, RD, SNM.
Funding acquisition: AB, ANGB, DM, GM, IJM, LAB, LM, RD, RR, RT.
Methodology: AL, ANGB, AW, CDH, IJM, JE, JH, KZ, LAB, RB, RR, SNM.
Project administration: AB, IJM, RT, RR.
Software: AL, AW, BA, BB, BV, CDH, CH, CM, DR, ET, IH, JC, JE, JH, JN, KZ, MB, MK, NC, PB, PF, RB, RF, RR, QX, SB, SM, SNM, TP, VM.
Supervision: AB, GM, IH, IJM, RT, RR, SNM.
Validation: BB, CMP, GM, MK, RF, SNM.
Visualization: AW, NC, SNM.
Writing (original draft): AB, ANGB, GM, IJM, LM, RD, RR, RT.
Writing (review and editing): AB, ANGB, AL, AW, GM, IJM, JW, LM, NC, RD, RR, RT, SNM.
All authors read and approved the final manuscript.

\section*{Competing Interests}

The author(s) declare that they have no competing interests.

\section*{Funding}

The work was funded by the
Science and Technology Facilities Council under grant number ST/V006126/1,
Biotechnology and Biological Sciences Research Council (grants BB/M003949/1, BB/R012679/1 and BB/S001034/1),
Engineering and Physical Sciences Research Council (grants EP/T004878/1 and EP/V054236/1),
Medical Research Council (grants MC\_UU\_00022/2 and MR/R00241X),
Natural Environment Research Council (grants NE/T004193/1 and NE/T010355/1),
The Scottish Government Rural and Environment Science and Analytical Services Division	(grants ``Centre of Expertise in Animal Disease Outbreaks'' and ``Strategic Research Programme''),
Scottish Government Chief Scientist Office (grant SPHSU17),
the UK Atomic Energy Authority, supported by BEIS,
the French National Research Agency (ANR) (IDEXLYON project, grant number ANR-16-IDEX-0005), and
Boehringer Ingelheim Animal Health France (The Veterinary Public Health (VPH) hub).

\section*{Acknowledgements}

This work was initially undertaken as a contribution to the Rapid Assistance in Modelling the Pandemic (RAMP) initiative, coordinated by the Royal Society. We are very grateful to all of the volunteers in the Scottish COVID-19 Response Consortium formed as part of that initiative -- \url{https://www.gla.ac.uk/scrc} -- who contributed to this project, and in particular to the team at Man Group for their assistance in designing and helping to build the initial version of this pipeline through the Royal Society's RAMP initiative, and the team at Invenia Labs -- Eric Perim Martins, Bella Wu, Sean Lovett and Alex Robson -- for their invaluable assistance with the Julia implementation.


\bibliography{arxiv.bib}

\begin{thebibliography}{10}

\bibitem{centre_for_mathematical_modelling_of_infectious_diseases_covid-uk_2021}
{Centre for Mathematical Modelling of Infectious Diseases}. covid-uk. London
  School of Hygiene \& Tropical Medicine; 2021.
\newblock Original-date: 2020-05-04T16:42:32Z.
\newblock Available from: \url{https://github.com/cmmid/covid-uk}.

\bibitem{mrc_centre_for_global_infectious_disease_analysis_covidsim_2021}
{MRC Centre for Global Infectious Disease Analysis}. {CovidSim}. MRC Centre for
  Global Infectious Disease Analysis; 2021.
\newblock Original-date: 2020-05-04T16:42:32Z.
\newblock Available from: \url{https://github.com/mrc-ide/covid-sim}.

\bibitem{adam_kucharski_2020-cov-tracing_2021}
{Adam Kucharski}. 2020-cov-tracing. London School of Hygiene \& Tropical
  Medicine; 2021.
\newblock Original-date: 2020-05-04T16:42:32Z.
\newblock Available from: \url{https://github.com/adamkucharski/2020-ncov}.

\bibitem{adam_kucharski_2020-ncov_2021}
{Adam Kucharski}. 2020-ncov. London School of Hygiene \& Tropical Medicine;
  2021.
\newblock Original-date: 2020-05-04T16:42:32Z.
\newblock Available from: \url{https://github.com/adamkucharski/2020-ncov}.

\bibitem{kucharski_early_2020}
Kucharski AJ, Russell TW, Diamond C, Liu Y, Edmunds J, Funk S, et~al.
\newblock Early dynamics of transmission and control of {COVID}-19: a
  mathematical modelling study.
\newblock The Lancet Infectious Diseases. 2020 May;20(5):553-8.
\newblock Available from:
  \url{https://linkinghub.elsevier.com/retrieve/pii/S1473309920301444}.

\bibitem{giordano_modelling_2020}
Giordano G, Blanchini F, Bruno R, Colaneri P, Di~Filippo A, Di~Matteo A, et~al.
\newblock Modelling the {COVID}-19 epidemic and implementation of
  population-wide interventions in {Italy}.
\newblock Nature Medicine. 2020 Jun;26(6):855-60.
\newblock Available from:
  \url{http://www.nature.com/articles/s41591-020-0883-7}.

\bibitem{kucharski_effectiveness_2020}
Kucharski AJ, Klepac P, Conlan AJK, Kissler SM, Tang ML, Fry H, et~al.
\newblock Effectiveness of isolation, testing, contact tracing, and physical
  distancing on reducing transmission of {SARS}-{CoV}-2 in different settings:
  a mathematical modelling study.
\newblock The Lancet Infectious Diseases. 2020 Oct;20(10):1151-60.
\newblock Available from:
  \url{https://linkinghub.elsevier.com/retrieve/pii/S1473309920304576}.

\bibitem{davies_association_2021}
Davies NG, Barnard RC, Jarvis CI, Russell TW, Semple MG, Jit M, et~al.
\newblock Association of tiered restrictions and a second lockdown with
  {COVID}-19 deaths and hospital admissions in {England}: a modelling study.
\newblock The Lancet Infectious Diseases. 2021 Apr;21(4):482-92.
\newblock Available from:
  \url{https://linkinghub.elsevier.com/retrieve/pii/S1473309920309841}.

\bibitem{keeling_models_2005}
Keeling MJ.
\newblock Models of foot-and-mouth disease.
\newblock Proceedings of the Royal Society B: Biological Sciences. 2005
  Jun;272:1195-202.

\bibitem{matthews_neighbourhood_2003}
Matthews L, Haydon DT, Shaw DJ, Chase-Topping ME, Keeling MJ, Woolhouse MEJ.
\newblock Neighbourhood control policies and the spread of infectious diseases.
\newblock Proceedings of the Royal Society B: Biological Sciences. 2003
  Aug;270(1525):1659-66.

\bibitem{noauthor_osr_2020}
{OSR} {Statement} regarding transparency of data related to {COVID}-19; 2020.
\newblock Available from:
  \url{https://osr.statisticsauthority.gov.uk/news/osr-statement-regarding-transparency-of-data-related-to-covid-19/}.

\bibitem{wilkinson_fair_2016}
Wilkinson MD, Dumontier M, Aalbersberg IJ, Appleton G, Axton M, Baak A, et~al.
\newblock The {FAIR} {Guiding} {Principles} for scientific data management and
  stewardship.
\newblock Scientific Data. 2016 Mar;3(1):160018.
\newblock Number: 1 Publisher: Nature Publishing Group.
\newblock Available from: \url{https://www.nature.com/articles/sdata201618}.

\bibitem{lamprecht_towards_2020}
Lamprecht AL, Garcia L, Kuzak M, Martinez C, Arcila R, Martin Del~Pico E,
  et~al.
\newblock Towards {FAIR} principles for research software.
\newblock Data Science. 2020 Jun;3(1):37-59.
\newblock Available from:
  \url{https://www.medra.org/servlet/aliasResolver?alias=iospress&doi=10.3233/DS-190026}.

\bibitem{goble_fair_2020}
Goble C, Cohen-Boulakia S, Soiland-Reyes S, Garijo D, Gil Y, Crusoe MR, et~al.
\newblock {FAIR} {Computational} {Workflows}.
\newblock Data Intelligence. 2020 Jan;2(1-2):108-21.
\newblock Available from: \url{https://doi.org/10.1162/dint_a_00033}.

\bibitem{fairplus_fair_nodate}
{FAIRplus}. {FAIR} {Cookbook};.
\newblock Available from:
  \url{https://fairplus.github.io/the-fair-cookbook/content/home.html}.

\bibitem{jacobsen_generic_2020}
Jacobsen A, Kaliyaperumal R, da~Silva~Santos LOB, Mons B, Schultes E, Roos M,
  et~al.
\newblock A {Generic} {Workflow} for the {Data} {FAIRification} {Process}.
\newblock Data Intelligence. 2020 Jan;2(1-2):56-65.
\newblock Available from: \url{https://doi.org/10.1162/dint_a_00028}.

\bibitem{davidson_digital_1998}
Davidson LA, Douglas K.
\newblock Digital {Object} {Identifiers}: {Promise} and {Problems} for
  {Scholarly} {Publishing}.
\newblock The Journal of Electronic Publishing. 1998 Dec;4(2).
\newblock Available from:
  \url{http://hdl.handle.net/2027/spo.3336451.0004.203}.

\bibitem{noauthor_orcid_nodate}
{ORCID};.
\newblock Available from: \url{https://orcid.org}.

\bibitem{noauthor_json-ld_nodate}
{JSON}-{LD} - {JSON} for {Linking} {Data};.
\newblock Available from: \url{https://json-ld.org/}.

\bibitem{cox_ten_2021}
Cox SJD, Gonzalez-Beltran AN, Magagna B, Marinescu MC.
\newblock Ten simple rules for making a vocabulary {FAIR}.
\newblock PLOS Computational Biology. 2021 Jun;17(6):e1009041.
\newblock Available from:
  \url{https://journals.plos.org/ploscompbiol/article?id=10.1371/journal.pcbi.1009041}.

\bibitem{scottish_covid-19_response_consortium_scottish_2020}
{Scottish COVID-19 Response Consortium}. Scottish {COVID}-19 {Response}
  {Consortium}; 2020.
\newblock Available from: \url{https://www.gla.ac.uk/scrc}.

\bibitem{royal_society_rapid_2020}
{Royal Society}. Rapid {Assistance} in {Modelling} the {Pandemic}: {RAMP};
  2020.
\newblock Available from:
  \url{https://royalsociety.org/topics-policy/Health%20and%20wellbeing/ramp/}.

\bibitem{groth_prov-overview_2013}
Groth P, Moreau L. {PROV}-{Overview}; 2013.
\newblock Available from:
  \url{https://www.w3.org/TR/2013/NOTE-prov-overview-20130430/}.

\bibitem{albertoni_w3c-dcat_nodate}
Albertoni R, Browning D, Cox S, Gonzalez-Beltran A, Perego A, Winstanely P.
  {W3C}-{DCAT};.
\newblock Available from:
  \url{https://www.w3.org/TR/2020/REC-vocab-dcat-2-20200204/}.

\bibitem{european_organization_for_nuclear_research_zenodo_2013}
{European Organization For Nuclear Research}, {OpenAIRE}. Zenodo. CERN; 2013.
\newblock Available from: \url{https://www.zenodo.org/}.

\bibitem{figshare_figshare_2021}
{figshare}. figshare; 2021.
\newblock Available from: \url{https://figshare.com}.

\bibitem{wolstencroft_fairdomhub_2016}
Wolstencroft K, Krebs O, Snoep JL, Stanford NJ, Bacall F, Golebiewski M, et~al.
\newblock {FAIRDOMHub}: a repository and collaboration environment for sharing
  systems biology research.
\newblock Nucleic Acids Research. 2016 Nov;45(D1):D404-7.
\newblock \_eprint:
  https://academic.oup.com/nar/article-pdf/45/D1/D404/8846631/gkw1032.pdf.
\newblock Available from: \url{https://doi.org/10.1093/nar/gkw1032}.

\bibitem{wolstencroft_seek_2015}
Wolstencroft K, Owen S, Krebs O, Nguyen Q, Stanford NJ, Golebiewski M, et~al.
\newblock {SEEK}: a systems biology data and model management platform.
\newblock BMC Systems Biology. 2015 Jul;9(1):33.
\newblock Available from: \url{https://doi.org/10.1186/s12918-015-0174-y}.

\bibitem{noauthor_splitgraph_nodate}
Splitgraph;.
\newblock Available from: \url{https://www.splitgraph.com}.

\bibitem{noauthor_dolt_2021}
Dolt is {Git} for {Data}!. DoltHub; 2021.
\newblock Original-date: 2019-07-24T17:46:25Z.
\newblock Available from: \url{https://github.com/dolthub/dolt}.

\bibitem{noauthor_dolthub_nodate}
{DoltHub} {Home};.
\newblock Available from: \url{https://www.dolthub.com/}.

\bibitem{noauthor_dataworld_nodate}
data.world {\textbar} {The} {Cloud}-{Native} {Data} {Catalog};.
\newblock Available from: \url{https://data.world/}.

\bibitem{scottish_covid-19_response_consortium_data_2021}
{Scottish COVID-19 Response Consortium}. Data {Pipeline} {Use} {Cases}; 2021.
\newblock Available from:
  \url{https://www.fairdatapipeline.org/docs/use_cases/}.

\bibitem{freire_provenance_2008}
Freire J, Koop D, Santos E, Silva C.
\newblock Provenance for {Computational} {Tasks}: {A} {Survey}.
\newblock Computing in Science \& Engineering. 2008 May;10(3):11-21.
\newblock Available from: \url{http://ieeexplore.ieee.org/document/4488060/}.

\bibitem{afgan_galaxy_2018}
Afgan E, Baker D, Batut B, van den Beek M, Bouvier D, Čech M, et~al.
\newblock The {Galaxy} platform for accessible, reproducible and collaborative
  biomedical analyses: 2018 update.
\newblock Nucleic Acids Research. 2018 Jul;46(W1):W537-44.
\newblock Available from:
  \url{https://academic.oup.com/nar/article/46/W1/W537/5001157}.

\bibitem{noauthor_apache_nodate}
Apache {Airflow};.
\newblock Available from: \url{https://airflow.apache.org}.

\bibitem{noauthor_kepler_nodate}
Kepler;.
\newblock Available from: \url{https://kepler-project.org/}.

\bibitem{noauthor_kaggle_nodate}
Kaggle: {Your} {Machine} {Learning} and {Data} {Science} {Community};.
\newblock Available from: \url{https://www.kaggle.com/}.

\bibitem{noauthor_jupyter_nodate}
Jupyter;.
\newblock Available from: \url{https://jupyter.org/index.html}.

\bibitem{information_commissioners_office_guide_2021}
{Information Commissioner's Office}. Guide to the {UK} {General} {Data}
  {Protection} {Regulation} ({UK} {GDPR}); 2021.
\newblock Available from:
  \url{https://ico.org.uk/for-organisations/guide-to-data-protection/guide-to-the-general-data-protection-regulation-gdpr/}.

\bibitem{noauthor_quilt_nodate}
Quilt;.
\newblock Available from: \url{https://github.com/quiltdata/quilt}.

\bibitem{noauthor_docker_nodate}
Docker;.
\newblock Available from: \url{https://www.docker.com}.

\bibitem{noauthor_quiltdata_nodate}
{QuiltData};.
\newblock Available from: \url{https://quiltdata.com}.

\bibitem{noauthor_floydhub_nodate}
{FloydHub} {Blog};.
\newblock Available from: \url{https://blog.floydhub.com/}.

\bibitem{noauthor_covid_2020}
Covid {Model}-{Runner}. COVID-19 Modeling; 2020.
\newblock Original-date: 2020-05-04T16:42:32Z.
\newblock Available from: \url{https://github.com/covid-modeling/model-runner}.

\bibitem{williamson_factors_2020}
Williamson EJ, Walker AJ, Bhaskaran K, Bacon S, Bates C, Morton CE, et~al.
\newblock Factors associated with {COVID}-19-related death using {OpenSAFELY}.
\newblock Nature. 2020 Aug;584(7821):430-6.
\newblock Available from:
  \url{http://www.nature.com/articles/s41586-020-2521-4}.

\bibitem{noauthor_opensafely_nodate}
{OpenSAFELY}: {Home};.
\newblock Available from: \url{https://www.opensafely.org/}.

\bibitem{davidson_provenance_2008}
Davidson SB, Freire J.
\newblock Provenance and {Scientific} {Workflows}: {Challenges} and
  {Opportunities}.
\newblock In: {SIGMOD} '08: {Proceedings} of the 2008 {ACM} {SIGMOD}
  international conference on {Management} of data. Vancouver, Canada:
  Association for Computing Machinery, New York, NY, United States of America;
  2008. p. 1345-50.

\bibitem{pasquier_practical_2017}
Pasquier T, Han X, Goldstein M, Moyer T, Eyers D, Seltzer M, et~al.
\newblock Practical {Whole}-{System} {Provenance} {Capture}.
\newblock In: Symposium on {Cloud} {Computing} ({SoCC}’17). ACM; 2017. .

\bibitem{pohly_hi-fi_2012}
Pohly D, Mclaughlin S, McDaniel P, Butler K.
\newblock Hi-{Fi}: {Collecting} high-fidelity whole-system provenance; 2012. p.
  259-68.

\bibitem{lerner_using_2018}
Lerner B, Boose E, Perez L.
\newblock Using {Introspection} to {Collect} {Provenance} in {R}.
\newblock Informatics. 2018;5(12).

\bibitem{jones_recordr_nodate}
Jones MB, Slaughter P, Jones C. recordr;.
\newblock Available from: \url{http://github.com/NCEAS/recordr}.

\bibitem{jackson_recipy_2018}
Jackson M, Wilson R, van~der Zwaan J, Steinbrook DW, Rathgeber F, Alegre R,
  et~al.. recipy; 2018.
\newblock Available from: \url{https://github.com/recipy/recipy}.

\bibitem{noauthor_core_nodate}
Core {Provenance} {Library};.
\newblock Available from: \url{https://github.com/ProvTools/prov-cpl}.

\bibitem{noauthor_git_nodate}
Git;.
\newblock Available from: \url{https://git-scm.com/}.

\bibitem{noauthor_git-annex_nodate}
git-annex;.
\newblock Available from: \url{https://git-annex.branchable.com/}.

\bibitem{noauthor_qri_nodate}
Qri;.
\newblock Available from: \url{https://qri.io/}.

\bibitem{noauthor_pachyderm_nodate}
Pachyderm;.
\newblock Available from: \url{https://www.pachyderm.com/}.

\bibitem{noauthor_kubernetes_2021}
Kubernetes; 2021.
\newblock Available from: \url{https://kubernetes.io}.

\bibitem{noauthor_dvc_nodate}
{DVC};.
\newblock Available from: \url{https://dvc.org/}.

\bibitem{noauthor_dagshub_nodate}
{DAGsHub};.
\newblock Available from: \url{https://dagshub.com/}.

\bibitem{noauthor_mlflow_nodate}
{MLflow};.
\newblock Available from: \url{https://mlflow.org}.

\bibitem{halchenko_datalad_2021}
Halchenko YO, Meyer K, Poldrack B, Solanky DS, Wagner AS, Gors J, et~al.
\newblock {DataLad}: distributed system for joint management of code, data, and
  their relationship.
\newblock Journal of Open Source Software. 2021 Jul;6(63):3262.
\newblock Available from:
  \url{https://joss.theoj.org/papers/10.21105/joss.03262}.

\bibitem{sefton_ro-crate_2021}
Sefton P, Carragáin E, Soiland-Reyes S, Corcho O, Garijo D, Palma R, et~al.
\newblock {RO}-{Crate} {Metadata} {Specification} 1.1.1.
\newblock NA. 2021 Feb.
\newblock Publisher: Zenodo.
\newblock Available from: \url{https://zenodo.org/record/4541002}.

\bibitem{scottish_covid-19_response_consortium_software-checklist_2020}
{Scottish COVID-19 Response Consortium}. software-checklist; 2020.
\newblock Available from:
  \url{https://github.com/ScottishCovidResponse/modelling-software-checklist}.

\bibitem{scottish_covid-19_response_consortium_fairdatapipeline_2021}
{Scottish COVID-19 Response Consortium}. {FAIRDataPipeline} {GitHub}
  {Organisation}; 2021.
\newblock Available from: \url{https://github.com/FAIRDataPipeline}.

\bibitem{blackwell_fair_2021}
Blackwell R, Brett A, Cook J, Cummings N, Field R, Gonzalez-Beltran A, et~al..
  The {FAIR} {Data} {Registry}; 2021.
\newblock Publisher: Zenodo.
\newblock Available from: \url{https://doi.org/10.5281/zenodo.5562749}.

\bibitem{zarebski_fair_2021}
Zarebski K, Reeve R, Reddyhoff D, Cummings N. The {FAIR} {Data} {Pipeline}
  command line tool; 2021.
\newblock Publisher: Zenodo.
\newblock Available from: \url{https://doi.org/10.5281/zenodo.5552779}.

\bibitem{mitchell_rdatapipeline_2021}
Mitchell S, Field R. {rDataPipeline} - {FAIR} {Data} {Pipeline} in {R}; 2021.
\newblock Publisher: Zenodo.
\newblock Available from: \url{https://doi.org/10.5281/zenodo.5338588}.

\bibitem{mitchell_datapipelinejl_2021}
Mitchell S, Burke M, Reeve R. {DataPipeline}.jl - {FAIR} {Data} {Pipeline} in
  {Julia}; 2021.
\newblock Publisher: Zenodo.
\newblock Available from: \url{https://doi.org/10.5281/zenodo.5270282}.

\bibitem{boskamp_javadatapipeline_2021}
Boskamp B. {javaDataPipeline} - {FAIR} {Data} {Pipeline} in {Java}; 2021.
\newblock Publisher: Zenodo.
\newblock Available from: \url{https://doi.org/10.5281/zenodo.5547492}.

\bibitem{field_pydatapipeline_2021}
Field R, Reddyhoff D. {pyDataPipeline} - {FAIR} {Data} {Pipeline} in {Python};
  2021.
\newblock Publisher: Zenodo.
\newblock Available from: \url{https://doi.org/10.5281/zenodo.5548002}.

\bibitem{field_c_2022}
Field R, Zarebski K. C++ {Implementation} of the {API} for the {FAIR} {Data}
  {Pipeline}.. Zenodo; 2022.
\newblock Available from: \url{https://zenodo.org/record/5877992}.

\bibitem{bjornstad_seirs_2020}
Bjørnstad ON, Shea K, Krzywinski M, Altman N.
\newblock The {SEIRS} model for infectious disease dynamics.
\newblock Nature Methods. 2020 Jun;17(6):557-8.
\newblock Available from:
  \url{http://www.nature.com/articles/s41592-020-0856-2}.

\bibitem{mitchell_rsimplemodel_2021}
Mitchell S. {rSimpleModel}; 2021.
\newblock Available from:
  \url{https://github.com/FAIRDataPipeline/rSimpleModel}.

\bibitem{mitchell_seirs_2021}
Mitchell S. {SEIRS} {Model} {Example}; 2021.
\newblock Available from:
  \url{https://www.fairdatapipeline.org/rSimpleModel/articles/SEIRS.html}.

\bibitem{boskamp_javasimplemodel_2021}
Boskamp B. {javaSimpleModel}; 2021.
\newblock Available from:
  \url{https://github.com/FAIRDataPipeline/javaSimpleModel}.

\bibitem{mitchell_seinrd_2021}
Mitchell S. {SEINRD} {Model} {Example}; 2021.
\newblock Available from:
  \url{https://www.fairdatapipeline.org/rSimpleModel/articles/SEINRD.html}.

\bibitem{offices_of_the_nuffield_professor_of_medicine_covid-19_2021}
{Offices of the Nuffield Professor of Medicine}. {COVID}-19 {Infection}
  {Survey}; 2021.
\newblock Available from:
  \url{https://www.ndm.ox.ac.uk/covid-19/covid-19-infection-survey}.

\bibitem{pooley_estimation_2021}
Pooley CM, Doeschl-Wilson AB, Marion G.
\newblock Estimation of age-stratified contact rates during the {COVID}-19
  pandemic using a novel inference algorithm.
\newblock Philosophical Transactions of the Royal Society A, special issue:
  "Technical challenges of modelling real-life epidemics and examples of
  overcoming these". in press.

\bibitem{european_commission_directorate_general_for_research_and_innovation_cost-benefit_2018}
{European Commission  Directorate General for Research and Innovation }, {PwC
  EU Services }.
\newblock Cost-benefit analysis for {FAIR} research data: cost of not having
  {FAIR} research data.
\newblock LU: Publications Office; 2018.
\newblock Available from: \url{https://data.europa.eu/doi/10.2777/02999}.

\end{thebibliography}
\bibliographystyle{vancouver} 

\newgeometry{total={18.5cm, 26.5cm}}

\section*{Supplementary Materials}

\setcounter{figure}{0}
\renewcommand{\figurename}{Supplementary Figure}

\begin{figure}[ht]
    \centering
    \includegraphics[width=0.8\textwidth]{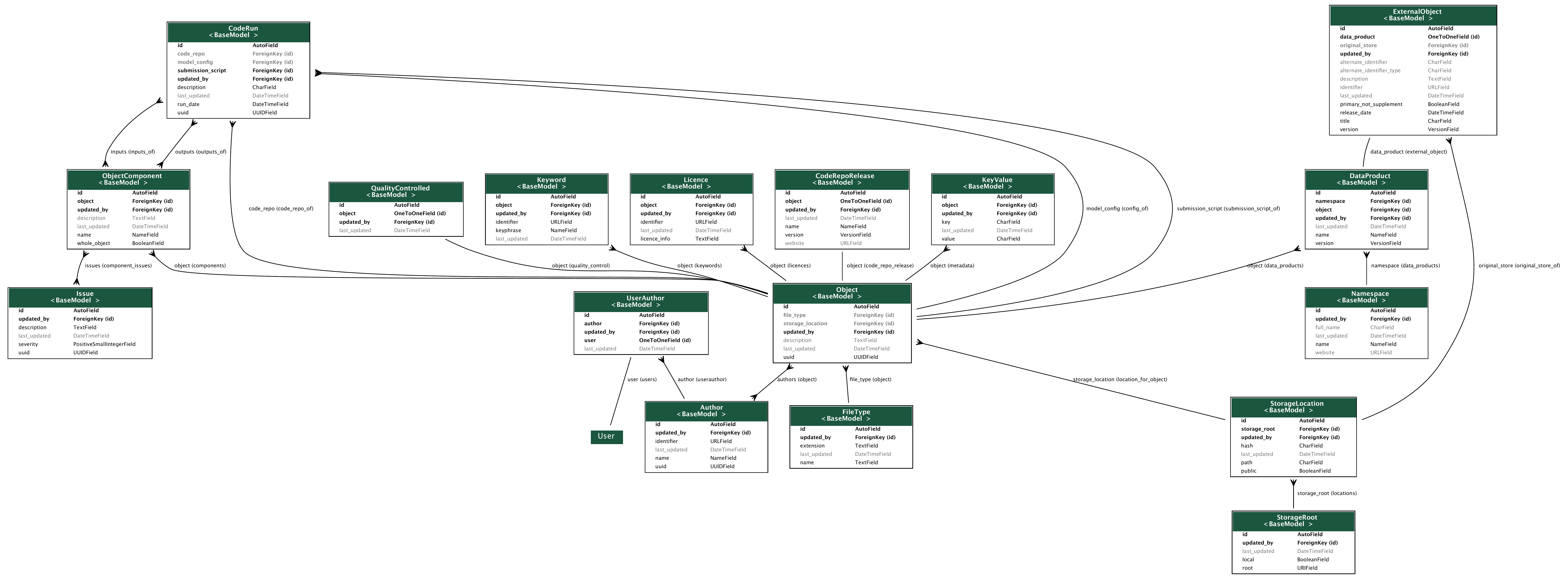}
    \caption{\textbf{Data Registry schema}}
\end{figure}

\begin{figure}[ht]
    \centering
    \includegraphics[width=0.8\textwidth]{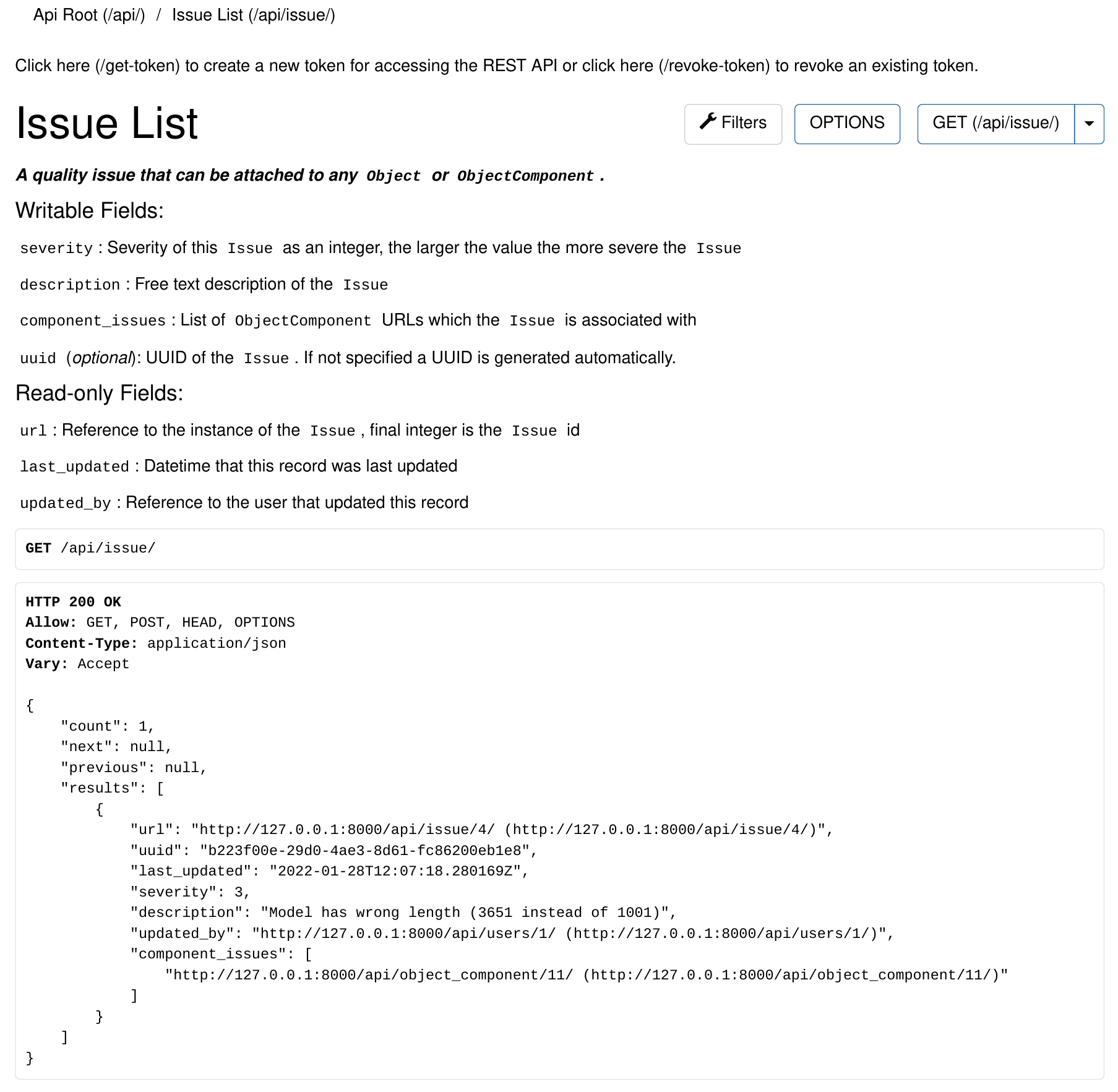}
    \caption{\textbf{Issue raised in the provenance of a model comparison}}
\end{figure}

\end{document}